\documentclass[twocolumn,aps,pra,longbibliography,superscriptaddress,nofootinbib,10pt]{revtex4-2}

\usepackage{bm}
\usepackage{bbm}
\usepackage[retainorgcmds]{IEEEtrantools}
\usepackage{graphicx}
\usepackage{mathrsfs}
\usepackage{amsmath}
\usepackage{amsfonts}
\usepackage{amssymb}
\usepackage{color,xcolor}
\usepackage{times}
\usepackage{braket}
\usepackage{nicefrac}
\usepackage{ragged2e}
\usepackage{braket}
\usepackage{tikz}

\usepackage[colorlinks=true,linkcolor=blue,urlcolor=blue,citecolor=blue,pdfusetitle]{hyperref}

\usepackage{blindtext}
\usepackage{stmaryrd} 

\newcommand{\tr}{\mathrm{Tr}}

\newtheorem{definition}{Definition}

\begin{document}

\title{Deterministic Equations for Feedback Control of Open Quantum Systems II:\\ Properties of the memory function}
\date{\today}
\author{Alberto J. B. Rosal}
\email{abezerra@ur.rochester.edu}
\affiliation{Department of Physics and Astronomy, University of Rochester, Rochester, New York 14627, USA}
\affiliation{University of Rochester Center for Coherence and Quantum Science, Rochester, New York 14627, USA}
\author{Patrick P. Potts}
\affiliation{Department of Physics and Swiss Nanoscience Institute,
University of Basel, Klingelbergstrasse 82 CH-4056, Switzerland}
\author{Gabriel T. Landi}
\affiliation{Department of Physics and Astronomy, University of Rochester, Rochester, New York 14627, USA}
\affiliation{University of Rochester Center for Coherence and Quantum Science, Rochester, New York 14627, USA}

\begin{abstract}
Feedback uses past detection outcomes to dynamically modify a quantum system and is central to quantum control. These outcomes can be stored in a memory, defined as a stochastic function of past measurements. 
In this work, we investigate the main properties of a general memory function subject to arbitrary feedback dynamics. 
We show that the memory can be treated as a classical system coupled to the monitored quantum system, and that their joint evolution is described by a hybrid bipartite state. This framework allows us to introduce information-theoretic measures that quantify the correlations between the system and the memory. Furthermore, we develop a general framework to characterize the statistics of the memory -- such as moments, cumulants, and correlation functions -- which can be applied both to general feedback-control protocols and to monitored systems without feedback.
As an application, we analyze feedback schemes based on detection events in a two-level system coupled to a thermal bath, focusing on protocols that stabilize either the excited-state population or Rabi oscillations against thermal dissipation.
\end{abstract}

\maketitle{}

\section{Introduction}
\label{sec:int}

Quantum feedback control has emerged as a powerful tool across a wide range of quantum technologies. It has been successfully applied to tasks such as cooling protocols \cite{CoolingGui,PhysRevA.107.023516,PhysRevLett.122.070603,PhysRevLett.123.223602,PhysRevLett.117.163601,PhysRevA.91.043812,PhysRevLett.96.043003,PhysRevA.74.012322,PhysRevLett.90.043001,PhysRevB.68.235328,PhysRevLett.119.123603,PhysRevA.98.023828}, entanglement generation \cite{Riste2013}, quantum error correction \cite{QuantumThermodynamicsForQuantumComputing,QECwithFeedback}, thermodynamic control \cite{DiscreteFeedbackExp,PatrickFluctuationTheo,Pekola2015,Prech2025QuantumThermodynamicsContinuousFeedback}, transport phenomena \cite{RevModPhys.75.1}, quantum control strategies \cite{DiscreteFeedback3,DiscreteFeedback2,Vijay2012, Minev2019CatchingReverseQuantumJump, Sayrin2011RealTimeFeedback,PhysRevA.67.052101,PhysRevLett.120.073601}, and quantum battery charging \cite{Mitchison2021chargingquantum}.
For thermodynamics, feedback has been experimentally explored in the realization of Maxwell’s demon in the quantum regime~\cite{QuantumDemon3,QuantumDemon2,QuantumDemon1}, motivating the development of generalized second laws of thermodynamics that explicitly incorporate measurement-based feedback processes~\cite{SecondLawFeedback3,SecondLawFeedback2,SecondLawFeedback1,PhysRevLett.133.140401}.
These advances underscore the fundamental importance of feedback in both quantum theory and experiment, and continue to drive the development of new quantum control strategies, playing a key role in the implementation of coherent quantum devices.

The idea behind feedback is that one can change the system based on previous outcomes. 
These iterative processes can be described by stochastic frameworks, but they usually require expensive computational resources and lack analytical insight, as they do not provide expressions for key quantities of the system -- such as the steady state, when it exists. 
On the other hand, Ref.~\cite{Rosal2025DeterministicFeedback} introduced a powerful deterministic framework capable of describing arbitrary feedback protocols. 
It provides analytical results in regimes inaccessible to previous approaches, and offers a unified structure that not only recovers known deterministic results~\cite{PatrickQFPME, WisemanMilburn1993_Homodyne, Wiseman1994_Feedback, FPTLandi} but also encompasses both continuous and discrete feedback schemes, as experimentally implemented in~\cite{DiscreteFeedback3, Minev2019CatchingReverseQuantumJump}.

For instance, consider that we can detect some observable of a system. 
This measurement process corresponds to the stage in which information about the system is acquired. 
After $n$ detections, one obtains a dataset $x_{1:n} \equiv (x_1, \ldots, x_n)$. 
We can store this information in a \emph{memory function}, denoted by $y_n$, which can then be used to feed an action back onto the system. 
The memory is thus a function of the dataset, $y_n = y_n(x_{1:n})$, representing the processed information used in the feedback action. 
It may correspond to the full dataset or a compressed version of it. 
For example, one may base the feedback action on the full record, $y_n = x_{1:n}$, or on a sample average, $y_n = \frac{1}{n}\sum_{i=1}^n x_i$. 
Since the outcomes $x_i$ are random, the memory $y_n$ is also a random variable.

In this work, we focus on the properties of a general memory function under an arbitrary feedback protocol. 
Our goal is to characterize its evolution by analyzing the corresponding probability distribution -- including its moments, cumulants, and time-correlation functions -- and to investigate the role played by the memory function in the feedback dynamics.
These properties are essential not only for understanding the influence of feedback on the memory's statistics, but also for broader applications in the theory of continuous quantum measurement, with potential implications for quantum metrology and parameter estimation.
Furthermore, this problem has never been systematically addressed, mainly due to the lack of general deterministic approaches for feedback control. 
Previous deterministic formulations~\cite{PatrickQFPME,yxqg-93jz,Wiseman1994_Feedback,WisemanMilburn1993_Homodyne,FPTLandi} were limited to specific choices of memory and measurement schemes and were defined only in the continuous-monitoring limit. As a consequence, they could not describe discrete feedback, despite its experimental relevance~\cite{DiscreteFeedback2,DiscreteFeedback3,DiscreteFeedbackExp}.

This work is organized as follows.  
In Sec.~\ref{sec:formalism}, we present the formalism used in this work, where Sec.~\ref{subsec:memory} introduces the notions of data processing and memory function, Sec.~\ref{subsec:fb_dyn} briefly presents the general framework developed in Ref.~\cite{Rosal2025DeterministicFeedback} to describe feedback dynamics, and Sec.~\ref{subsec:cont_mon_limit} defines the feedback dynamics in the continuous-monitoring limit.
In Sec.~\ref{subsec: Hybrid states}, we demonstrate that a general feedback dynamics admits a hybrid quantum--classical description, thereby allowing for the consistent inclusion of informational measures to quantify system--memory correlations.
In Sec.~\ref{sec:FCS_definitions}, we present a detailed analysis of the statistical properties of $y_n$ under a general feedback dynamics, including its moments, cumulants, time correlations, and other related quantities.
Sec.~\ref{sec: no feedback and indirect feedback} addresses the no-feedback scenario, as well as the case of indirect feedback, and Sec.~\ref{sec: examples of memory functions} presents explicit examples of memory functions commonly used in feedback strategies.
Finally, in Sec.~\ref{sec:application}, we apply our formalism to two feedback protocols: the first stabilizes the excited state of a qubit against thermal interaction based on quantum jump detections, and the second stabilizes the Rabi oscillations of a qubit coupled to a thermal bath using projective measurements, similar to the protocol experimentally implemented in Ref.~\cite{DiscreteFeedback3}.

\section{Formalism}
\label{sec:formalism}
\subsection{Data processing and memory}
\label{subsec:memory}
Let us consider the measurement of a physical observable at discrete times $t_i$ ($i = 1, 2, \ldots$), where $x_i$ denotes the outcome obtained at time $t_i$.  
After $n$ measurements, the accumulated data are represented by $x_{1:n} \equiv (x_1, \ldots, x_n)$.  
We define the \textit{stochastic memory} $y_n$ as a general function of this data, $y_n = y_n(x_{1:n})$, which serves as a central quantity in this work.  
The memory represents a processed version of the original dataset $x_{1:n}$ and may correspond either to a compressed summary of the detected outcomes or to the complete measurement history.
Furthermore, a memory that satisfies an update rule of the form
\begin{equation}
    y_{n} = f_n(x_n,y_{n-1})~
\end{equation}
for some function $f_n$ is referred to as \emph{causal memory}. 

It is important to emphasize the generality of the class of causal memories.  
For instance, if we define $y_n = \text{append}(y_{n-1}, x_n)$, the memory corresponds to the full dataset, $y_n = x_{1:n} = (x_1, \ldots, x_n)$.  
This represents a causal memory with no compression: the processed data $y_n$ coincides with the original dataset $x_{1:n}$.  
In practice, however, the feedback action often does not require the entire detection history, and $y_n$ may serve as a compressed representation of the data.  
For example, one may define the memory as the most recent outcome, $y_n = x_n$, which can be written as $y_n = f_n(x_n, y_{n-1})$ with an update function $f_n(x, y) = x$.  
This case corresponds to a maximal compression, where all previous outcomes are discarded and only the current detection is retained.
Moreover, the memory can be defined as the running average of the outcomes, 
$y_n = \frac{1}{n}\sum_{i=1}^{n} x_i$, which satisfies the recursive relation 
$y_n = \big(x_n + (n-1)y_{n-1}\big)/n$. 
Hence, the corresponding update function is $f_n(x, y) = \big(x + (n-1)y\big)/n$. 
In this case, the full dataset is taken into account, but its information is compressed into the average $\frac{1}{n}\sum_{i=1}^{n} x_i$.

The key point is that the stochastic memory $y_n$ encodes the relevant information extracted from the stochastic data $x_{1:n}$ and can therefore be used to construct feedback strategies.  
The choice of memory is highly context dependent and determined by the specific task at hand.
By defining an appropriate update function $f_n$, one can describe a broad class of memories, encompassing both linear cases -- where $y_n$ is a linear function of the outcomes $x_i$ -- and nonlinear ones.
Further discussions and examples of memory functions are provided in the Supplemental Material of Ref.~\cite{Rosal2025DeterministicFeedback}.

The probability distribution of the stochastic memory, $P(y_n = y)$, provides the statistics of the processed outcomes.  
These memory statistics are experimentally accessible and establish a direct connection to measurable observables.  
In this work, we develop a systematic framework to characterize the statistical properties of a stochastic memory under a general feedback protocol. 
To this end, we employ the deterministic-equation formalism developed in Ref.~\cite{Rosal2025DeterministicFeedback}, which is briefly summarized below.

\subsection{Feedback dynamics}
\label{subsec:fb_dyn}

A \emph{sequential measurement scheme} consists of performing consecutive detections of a system observable.  
Let us consider a measurement described by a set of Kraus operators 
$\{V_x\}$, satisfying the completeness relation 
$\sum_x V_x^\dagger V_x = \mathbb{I}$, where $\mathbb{I}$ denotes the identity operator. These measurements are applied at times $t_n$ with $n = 1, 2, \ldots$, where $t_i < t_j$ for $i < j$.
Between detections, the system may evolve from $t_n$ to $t_{n+1}$ according to the dynamical map $\Lambda_{t_{n+1}, t_n}$.  
The map describing the overall transformation from $t_n$ to $t_{n+1}$ -- including both the dynamical evolution and the measurement process -- is then given by $\mathcal{M}_x(\rho) = V_x \, [\Lambda_{t_{n+1}, t_n}(\rho)] \, V_x^\dagger$.
These maps, also referred to as \emph{instruments}~\cite{Wilde2021}, are trace non-increasing and satisfy the condition that $\sum_x \mathcal{M}_x$ defines a quantum channel, i.e., a completely positive and trace-preserving map~\cite{PRXQuantum.2.030201}.
Let us suppose that the system is initially prepared in the state $\rho_0$.  
After $n$ measurements, one obtains a sequence of outcomes $x_{1:n} = (x_1, x_2, \ldots, x_n)$, and the corresponding conditional state of the system is denoted by $\rho_{x_{1:n}}$.
The representation of the dataset is given by the memory $y_n$, which is used in the feedback action to modify the measurement scheme (determined by the Kraus operators $\{V_x\}$) \textit{and/or} the system's dynamical evolution (described by the map $\Lambda_{t_{n+1}, t_n}$) at each step.

Thus, in general, a sequential measurement scheme can be described by applying the instruments $\mathcal{M}_x$ at each time step $t_n$.
A feedback mechanism corresponds to dynamically adjusting these instruments $\mathcal{M}_x$ at each measurement step based on the memory $y_n$.
The most general feedback mechanism is therefore implemented by allowing the instrument at time $t_{n+1}$ to depend on the memory $y_n$, i.e., $\mathcal{M}_{x_{n+1}}(y_n)$.
Therefore, the feedback dynamics is described by the following stochastic rules
\begin{equation}
\begin{aligned}
\label{eq: fb update rules}
    P(x_{n+1} | x_{1:n}) &= \mathrm{Tr}[\mathcal{M}_{x_{n+1}}(y_{n})\rho_{x_{1:n}}],
    \\[0.2cm]
    \rho_{x_{1:n+1}} &= \frac{\mathcal{M}_{x_{n+1}}(y_{n})\rho_{x_{1:n}}}{P(x_{n+1} | x_{1:n})},
\end{aligned}    
\end{equation}
where $P(x_{n+1}|x_{1:n})$ is the probability of obtaining the outcome $x_{n+1}$ given the previous dataset $x_{1:n}$.

In Ref.~\cite{Rosal2025DeterministicFeedback}, a deterministic equation was derived to describe general feedback protocols based on causal memories. For simplicity, let us assume that $y_n$ takes values in a discrete set (the continuous case will be discussed below). To introduce the deterministic equation governing the feedback dynamics, we define the memory-resolved state as
\begin{equation}
\label{eq: def. memory resolved state}
\varrho_n(y) \equiv E[\rho_{x_{1:n}} \delta_{y, y_n}]~,
\end{equation}
where $E[\cdot]$ denotes the average over all possible measurement outcomes $x_{1:n}$, and $\delta_{y, y_n}$ is the Kronecker delta. 
The trace of this state, $\mathrm{Tr}[\varrho_n(y)] = P(y_n = y)$, gives the probability distribution of the stochastic memory at time $t_n$, while summing over all possible memory values yields the unconditional system state, $\sum_y \varrho_n(y) = \bar{\rho}_n$.

For a given causal memory $y_n = f_n(x_n, y_{n-1})$ and a feedback protocol defined by the instruments $\{\mathcal{M}_x(y)\}$, it was shown in Ref.~\cite{Rosal2025DeterministicFeedback} that the memory-resolved state evolves from $t_n$ to $t_{n+1}$ according to
\begin{equation}
    \label{eq: deterministic eq 1}
    \varrho_{n+1}(y) = \sum_{x', y'} \delta_{y, f_{n+1}(x', y')} \mathcal{M}_{x'}(y') \varrho_n(y')~,
\end{equation}
where the sum runs over all possible outcomes $x'$ and memory values $y'$. Thus, Eq.~\eqref{eq: deterministic eq 1} describes the feedback dynamics at each step $t_n$.

In summary, although the feedback evolution is intrinsically stochastic [Eq.~\eqref{eq: fb update rules}], it can be fully captured by the deterministic evolution of the memory-resolved state $\varrho_t(y)$ [Eq.~\eqref{eq: deterministic eq 1}]. The framework can be summarized as follows:
\begin{definition}{A feedback protocol is defined in three steps: }
\label{def: definition of a fb protocol}
    \begin{enumerate}
    \item \textbf{Detections:} The sequential measurements performed at times $t_n$ are described by a set of super-operators $\{\mathcal{M}_x\}$, where the sum $\sum_x \mathcal{M}_x$ is completely positive and trace-preserving (CPTP).
    \item \textbf{Memory:} The stochastic memory $y_n = y_n(x_1, \ldots, x_n)$ is a function of the previous detection outcomes and satisfies the recursive relation $y_n = f_n(x_n, y_{n-1})$.
    \item \textbf{Feedback action:} We implement the feedback by allowing the instruments to depend on the memory, i.e., $\mathcal{M}_x \to \mathcal{M}_x(y)$.
\end{enumerate}
Then, the feedback dynamics is described by Eq.~\eqref{eq: deterministic eq 1}.
\end{definition}
Note that Eq.~\eqref{eq: fb update rules} can be used to numerically simulate realizations of a feedback protocol. However, such simulations are computationally expensive, as they require averaging over many trajectories to obtain meaningful statistics and smooth results. More importantly, they provide no analytical insight, and their outcomes are strongly parameter dependent.
In contrast, Eq.~\eqref{eq: deterministic eq 1} allows for analytical or simple numerical solutions. As shown in Ref.~\cite{Rosal2025DeterministicFeedback}, it also provides a systematic framework to obtain the feedback steady state and yields analytical results in regimes not accessible to previous feedback formulations.

\subsection{Continuous monitoring limit}
\label{subsec:cont_mon_limit}

Equation~\eqref{eq: deterministic eq 1} describes the evolution of the memory-resolved state from $t_n$ to $t_{n+1}$, assuming that $y_n$ is a discrete random variable, where the sum $\sum_{y'}$ runs over all its possible values.
We can also consider either time-continuous or continuous-valued cases.
The time-continuous case corresponds to the \emph{continuous monitoring limit}, while the continuous-valued case corresponds to a stochastic memory that takes values in a continuous set.
Equation~\eqref{eq: deterministic eq 1} can be straightforwardly adapted to both situations as follows.

Let us consider that measurements are performed at regular times $t_n = n ~\delta t$, with $n =  1, 2, \cdots$, and $\delta t \to 0$.
In this limit, the stochastic memory becomes a continuous-time stochastic process, denoted by $y_t$, and it represents a generic function of the previous outcomes $x_{t'}$ for any $t'\leq t$.
Note that $y_t$ may take values in either discrete or continuous sets.
The conditional state associated with the dataset $\{x_{t'}\}_{t'\leq t}$ is denoted by $\rho_t^c$, and for a discrete-valued $y_t$, the memory-resolved state [Eq.~\eqref{eq: def. memory resolved state}] at time $t$ becomes $\varrho_t(y) = E[\rho_t^c \delta_{y,y_t}]$. Hence, the probability distribution of $y_t$ is given by $P_t(y) = \tr[\varrho_t(y)]$, and the system's state at time $t$ is $\bar{\rho}_t = \sum_y~\varrho_t(y)$. Note that the normalization of $P_t(y)$ is satisfied, since $\sum_y P_t(y) = \tr[\sum_y \varrho_t(y)] = \tr[\bar{\rho}_t] = 1$ for any $t\geq0$.
For example, Ref.~\cite{Rosal2025DeterministicFeedback} developed a general feedback scheme based on the continuous monitoring of quantum jumps, in which a discrete-valued memory records the last detected jump channel.

For a continuous-valued memory, the memory-resolved state is defined as $\varrho_t(y) = E[\rho_t^c \delta(y -y_t)]$, where $\delta(y-y_t)$ is the Dirac delta. In this continuous case, $P_t(y) = \tr[\varrho_t(y)]$ becomes a probability density function, and the unconditional state of the system is given by $\bar{\rho}_t = \int dy~\varrho_t(y)$. The normalization of $P_t(y)$ becomes $\int dy~P_t(y) = \tr[\int dy~ \varrho_t(y)] = \tr[\bar{\rho}_t] =1 $. 
For instance, Ref.~\cite{PatrickQFPME} developed a feedback scheme based on the low-pass filter $y_t = \int_0^t ds\, \gamma e^{-\gamma (t-s)} x_s$, which corresponds to a time-continuous and continuous-valued memory.

By definition, \emph{discrete feedback} refers to protocols based on sequential measurements performed at times $t_n = n~\delta t$, with a finite $\delta t>0$. 
In contrast, \emph{continuous feedback} relies on continuous monitoring, defined in the limit $\delta t \rightarrow 0$. 
Equation~\eqref{eq: deterministic eq 1} encompasses both cases: in the continuous limit, it reduces to an integral or differential equation~\cite{Rosal2025DeterministicFeedback}.
Consequently, one can move from the discrete to the continuous description by replacing sums with integrals and Kronecker deltas with Dirac deltas.
In contrast, previous deterministic formulations were limited to continuous monitoring under specific measurement schemes, such as homodyne detection~\cite{WisemanMilburn1993_Homodyne} or weak Gaussian measurements~\cite{PatrickQFPME}.

In the following, we focus on a discrete-valued memory, although all results extend straightforwardly to continuous-valued cases.
We denote the memory at time $t$ by $y_t$ and its value at discrete times $t_n$ by $y_n$, with the same notation applying to the memory-resolved states $\varrho_t(y)$ and $\varrho_n(y)$. 
Unless explicitly stated otherwise, all statements referring to the continuous-time memory $y_t$ also hold for the discrete-time memory $y_n$.

\section{Hybrid classical-quantum description}
\label{subsec: Hybrid states}

The memory-resolved state $\varrho_t(y)$ in Eq.~\eqref{eq: deterministic eq 1} describes both a classical degree of freedom, represented by the stochastic memory $y_t$ with probability distribution $P_t(y) = \tr[\varrho_t(y)]$, and a quantum degree of freedom, given by the quantum state $\bar{\rho}_t = \sum_y \varrho_t(y)$. 
In this section, we show that this hybrid classical-quantum system can be described using a joint Hilbert space $\mathcal{H} = \mathcal{H}_\text{s} \otimes \mathcal{H}_\text{cl}$, where $\mathcal{H}_\text{s}$ is the Hilbert space of the quantum system, and $\mathcal{H}_\text{cl}$ represents the Hilbert space of the memory $y_t$.

Let us consider the quantum component described by $\mathcal{H}_\text{s}$. We introduce the \emph{memory-conditioned} states $\rho_t(y)$ as
\begin{equation}
\label{eq: memory-conditioned states}
    \rho_t(y) \equiv \frac{\varrho_t(y)}{\tr[\varrho_t(y)]}~.
\end{equation}
Here, $\rho_t(y)$ represents the quantum state conditioned on the classical realization $y_t = y$ of the stochastic memory. In fact, since $P_t(y) = \tr[\varrho_t(y)]$ and $\bar{\rho}_t = \sum_y \varrho_t(y)$, it follows that
\begin{equation}
\label{eq: ensemble of the memory-conditioned states}
    \bar{\rho}_t = \sum_y P_t(y) \rho_t(y)~.
\end{equation}
In other words, the unconditional state $\bar{\rho}_t$ of the system can be interpreted as an ensemble $\{P_t(y), \rho_t(y)\}$ of the memory-conditioned states defined in Eq.~\eqref{eq: memory-conditioned states}.
On the other hand, the classical component is represented by the memory Hilbert space $\mathcal{H}_\text{cl}$, spanned by an orthonormal basis $\{\ket{y}\}_y$, where each $y$ corresponds to a possible realization of the stochastic memory $y_t$.
The classical state $\rho_t^{\mathrm{cl}}$ is then defined as
\begin{equation}
\label{eq: classical-state}
    \rho_t^{\mathrm{cl}} \equiv \sum_y P_t(y) \ket{y}\bra{y}~.
\end{equation}
The state $\rho_t^{\text{cl}}$ is referred to as a classical state because it is always diagonal in the memory basis $\ket{y}$ and thus represents the classical state of the memory.
 
The full classical–quantum system can now be described by introducing a joint state $\rho_{\text{sm}}(t)$ in the composite Hilbert space $\mathcal{H}_\text{s} \otimes \mathcal{H}_\text{cl}$, defined as
\begin{equation}
\label{eq: composed quantum-classical state}
\rho_{\text{sm}}(t) \equiv \sum_y P_t(y) \rho_t(y) \otimes \ket{y}\bra{y}~.
\end{equation}
The unconditional quantum state is obtained by tracing over the memory Hilbert space $\mathcal{H}_\text{cl}$,
\begin{equation}
\tr_\text{cl}[\rho_{\text{sm}}(t)] = \sum_y P_t(y) \rho_t(y) = \sum_y \varrho_t(y) = \bar{\rho}_t~,
\end{equation}
while the classical state is recovered by tracing out the quantum subsystem,
\begin{equation}
\label{eq: classical memory state}
\tr_\text{s}[\rho_{\text{sm}}(t)] = \sum_y P_t(y) \ket{y}\bra{y} = \rho_t^{\mathrm{cl}}~.
\end{equation}

In summary, a quantum system subjected to a sequential measurement scheme with data processing described by a memory $y_t$ can be viewed as comprising two distinct degrees of freedom: a classical one, corresponding to the memory, and a quantum one, associated with the system’s state. 
The combined classical–quantum system is described within the extended Hilbert space $\mathcal{H}_\text{s} \otimes \mathcal{H}_\text{cl}$, with the joint state $\rho_{\text{sm}}(t)$ defined in Eq.~\eqref{eq: composed quantum-classical state}.

The feedback dynamics, described by Eq.~\eqref{eq: deterministic eq 1}, can be reformulated in terms of a dynamical map acting on the joint state $\rho_{\text{sm}}(t)$. Let us consider sequential measurements applied at times $t_n$, described by the instruments $\mathcal{M}_{x}(y)$, as introduced in Def.~\ref{def: definition of a fb protocol}. Given a causal memory $y_n$ satisfying the relation $y_n = f_n(x_n, y_{n-1})$, we define the linear map $\Omega_n$ acting on $\mathcal{H}_\text{s} \otimes \mathcal{H}_\text{cl}$ as
\begin{equation}
\label{eq: propagator of the joint system-memory dynamics}
\Omega_n(\sigma \otimes \ket{y'}\bra{y'}) \equiv \sum_{x', y} \left( \mathcal{M}_{x'}(y') \sigma \otimes \ket{y}\bra{y} \right) \delta_{y, f_n(x', y')}~,
\end{equation}
where $\sigma$ is an arbitrary quantum state of the system. Applying this map to $\rho_{\text{sm}}(n) \equiv \rho_{\text{sm}}(t = t_n) $ and using Eq.~\eqref{eq: deterministic eq 1}, we obtain
\begin{equation}
\label{eq: evolution of the composed state}
    \rho_{\text{sm}}(n+1) = \Omega_n \rho_{\text{sm}}(n)~,
\end{equation}
where $\rho_{\text{sm}}(n+1)$ is the joint classical-quantum state defined in Eq.~\eqref{eq: composed quantum-classical state}, evaluated at time $t_{n+1}$.
Using the Kronecker delta to perform the sum over $y$ in Eq.~\eqref{eq: propagator of the joint system-memory dynamics}, one obtains
\begin{equation}
\label{eq: propagator of the joint system-memory evolution 2}
\Omega_n = \sum_{x',y'} \mathcal{M}_{x'}(y') \otimes \mathcal{J}_{f_n(x',y'),y'}~,
\end{equation}
where
\begin{equation}
\mathcal{J}_{y,y'}\bullet \equiv \ket{y}\bra{y'}~\bullet~\ket{y'}\bra{y}
\end{equation}
is the super-operator describing the memory update in the classical memory space $\mathcal{H}_{\text{cl}}$.

Therefore, the feedback dynamics originally expressed in terms of the memory-resolved state $\varrho_n(y)$ via Eq.~\eqref{eq: deterministic eq 1} can be equivalently reformulated as the evolution of the joint state $\rho_{\text{sm}}(n)$, governed by the dynamical map $\Omega_n$ on the composite Hilbert space $\mathcal{H}_\text{s} \otimes \mathcal{H}_\text{cl}$, as given by Eq.~\eqref{eq: evolution of the composed state}. 
Equation~\eqref{eq: propagator of the joint system-memory evolution 2} shows that the joint system-memory evolution under general feedback can be decomposed into two components: the system dynamics governed by $\mathcal{M}_{x'}(y')$ and the memory update $y' \to y = f(x',y')$, described by $\mathcal{J}_{f(x',y'),y'}$.

The advantage of using the composed state $\rho_{\text{sm}}(t)$ is that it corresponds to an authentic density matrix associated to the Hilbert space $\mathcal{H}_\text{s}\otimes\mathcal{H}_\text{cl}$, and one can use it to introduce informational quantities relating the system and the memory. 
For instance, the quantum-classical mutual information $\mathcal{I}_t(\text{s}:\text{m})$ between system and memory $y_t$ is defined as
\begin{equation}
\label{eq: mutual classical-quantum information}
    \mathcal{I}_t(\text{s}:\text{m})\equiv S(\bar{\rho}_t)+S(\rho_t^{\text{cl}}) - S(\rho_{\text{sm}}(t))~,
\end{equation}
where $S(\rho) = -\tr[\rho \log{\rho}]$ is the von Neumann entropy, and $\mathcal{I}_t(\text{s}:\text{m})$ quantifies the correlations between the memory $y_t$ and the quantum system. 

Equation~\eqref{eq: composed quantum-classical state} describes a \emph{separable state}~\cite{NielsenChuang2010,Wilde2021}, i.e., a convex linear combination of the tensor product $\rho_t(y)\otimes\ket{y}\bra{y}$, where $\sum_{y} P_t(y) = 1$ and $P_t(y)\geq 0$. 
As a result, the system is never entangled with the memory $y_t$, which is consistent with the nature of the memory function, corresponding to a classical stochastic process.
Furthermore, since the states $\{\ket{y}\}$ constitute an orthonormal basis for the memory space, the entropy of the joint state $\rho_{\text{sm}}(t)$ is given by
\begin{equation}
    S\!\left(\rho_{\text{sm}}(t)\right) 
    = H\!\left(P_t(y)\right) 
      + \sum_{y} P_t(y)\, S\!\left(\rho_t(y)\right),
\end{equation}
where $H(P_t(y)) = -\sum_{y} P_t(y)\log P_t(y)$ is the Shannon entropy of the classical memory distribution, and $S(\rho_t(y))$ is the von Neumann entropy of the memory-conditioned states $\rho_t(y)$.
Using that $P_t(y) = \tr[\varrho_t(y)]$ and $\rho_t(t) = \varrho_t(y) /P_t(y)$, one has
\begin{equation}
    S(\rho_{\text{sm}}(t)) = \sum_{y} S(\varrho_t(y)) = -\sum_y \tr\big[\varrho_t(y) \log{\varrho_t(y)}\big]~.
\end{equation}
Hence, the memory-resolved state $\varrho_t(y)$ also encodes the total von Neumann entropy of the hybrid classical--quantum system composed of the quantum system under detection and the memory function $y_t$, and described by the bipartite state $\rho_{\text{sm}}(t)$.

If one has a system's observable $\mathcal{O}$ (that, at this moment, we suppose to be independent of the memory $y_t$), the average is given by $\braket{\mathcal{O}}_t = \tr_s[\bar{\rho}_t \mathcal{O}] = \sum_y \tr_s[\mathcal{O} \varrho_t(y)]$. Then, we can rewrite it in terms of $\rho_{\text{sm}}(t)$ as
\begin{equation}
    \braket{\mathcal{O}}_t = \tr[(\mathcal{O}\otimes\mathbb{I}_\text{cl})\rho_{\text{sm}}(t)],
\end{equation}
where $\mathbb{I}_\text{cl}$ represents the identity operator in the Hilbert space $\mathcal{H}_\text{cl}$. Finally, the covariance between the memory and a given system observable $\mathcal{O}$ is defined as
\begin{equation}
\label{eq: def memory covariance}
\text{cov}(y_t,\mathcal{O}) \equiv \tr[(\mathcal{O}\otimes\mathrm{Y})\rho_{\text{sm}}(t)] - \braket{\mathcal{O}}_t\braket{\mathrm{Y}}_t~,
\end{equation}
where $\mathrm{Y}$ is the \textit{memory operator} given by
\begin{equation}
    \mathrm{Y} \equiv \sum_y y |y\rangle\langle y|~,
\end{equation}
and one has
\begin{equation}
    \braket{\mathrm{Y}}_t \equiv\tr[(\mathbb{I}_\text{S}\otimes \mathrm{Y})\rho_{\text{sm}}(t)] =\sum_y y~P_t(y) = \braket{y_t}~.
\end{equation}

Note that, since the feedback protocol is conditioned on previous measurement outcomes, the evolution of the quantum system generally depends on its past history. In this sense, the reduced dynamics of the system exhibits memory effects and may be regarded as non-Markovian. In our framework, we have shown that, for certain classes of memories, the joint system-memory state $\rho_{\mathrm{sm}}(t)$ is governed by a time-local Lindblad-like master equation~\cite{Rosal2025FCSwithFB}. Consequently, the joint evolution of the system and memory can be regarded as Markovian~\cite{Rivas2014NonMarkovianity}. However, after tracing out the memory, the reduced state $\bar{\rho}_t=\mathrm{Tr}[\rho_{\mathrm{sm}}(t)]$ is generally not described by a time-local Markovian master equation. This illustrates a situation in which the extended system-memory dynamics is Markovian, while the reduced dynamics of the quantum system is generally non-Markovian due to the memory retained in the feedback controller.

In what follows, we present a general framework to derive the statistical properties of an arbitrary memory $y_t$ from the memory-resolved state $\varrho_t(y)$, including characteristic functions, moments, cumulants, and time correlations.
These results directly connect the feedback dynamics to measurable quantities, such as the average energy or the mean values of a memory.
An equivalent analysis can also be performed using the joint state $\rho_{\text{sm}}(t)$ defined in Eq.~\eqref{eq: composed quantum-classical state}.

\section{Memory statistics}
\label{sec:FCS_definitions}

\subsection{Average quantities, memory moments and cumulants}
Suppose we have some physical quantity of the system that depends on the stochastic memory, denoted by $\mathcal{O}(y_t)$. Then, the average value $\left<\mathcal{O}(y_t)\right>$ is given by
\begin{equation}
\left<\mathcal{O}(y_t)\right> = \sum_y P_t(y) \left<\mathcal{O}\right>_y~,
\end{equation}
where $\left<\mathcal{O}\right>_y = \tr[\mathcal{O}(y) \rho_t(y)]$ is the average value of $\mathcal{O}$ given the realization $y_t = y$, and the sum runs over all possible values of $y_t$ at time $t$. If $y_t$ assumes values in a continuous set, we can replace the sum by an integral, as discussed before. By applying Eq.~\eqref{eq: memory-conditioned states} and using that $P_t(y) = \tr[\varrho_t(y)]$, one finds 
\begin{equation}
\label{eq: average values for FB dynamics}
  \left<\mathcal{O}(y_t)\right> = \sum_y \tr[\mathcal{O}(y) \varrho_t(y)]~.
\end{equation}
Note that if we remove the memory dependence on $\mathcal{O}$, we recover the well-known expression for averages, $ \left<\mathcal{O}\right> =  \tr[\mathcal{O} \bar{\rho}_t]$, where $\bar{\rho}_t = \sum_y \varrho_t(y)$. 
For instance, Eq.~\eqref{eq: average values for FB dynamics} can be used to compute the average energy of the system when the Hamiltonian depends on the memory, providing important insights into the thermodynamic properties of the feedback dynamics~\cite{Prech2025QuantumThermodynamicsContinuousFeedback}.

Given a memory $y_t$, one may be interested in its moments $\braket{y_t^j}$, which can be obtained from the \textit{characteristic function}. Let us introduce the Fourier transform of the memory-resolved state as
\begin{equation}
\label{eq: fourier transform of the memory-resolved state}
    \varrho_t(\chi) \equiv \sum_y e^{iy\chi} \varrho_t(y)~,
\end{equation}
where the continuous variable $\chi$ is called \textit{counting field} in the context of full counting statistics~\cite{Tutorial}. One can recover the memory-resolved state by applying the inverse Fourier transform,
\begin{equation}
\label{eq: inverse fourier transform of the memory-resolved state}
    \varrho_t(y) = \int \frac{d\chi}{2\pi} e^{-i y \chi} \varrho_t(\chi)~.
\end{equation}
In some cases, the Fourier transform of the memory-resolved state can simplify the deterministic equation governing the feedback dynamics. In such cases, one may solve the equation for $\varrho_t(\chi)$ and subsequently recover $\varrho_t(y)$ by applying the inverse Fourier transform~\eqref{eq: inverse fourier transform of the memory-resolved state}.

Note that $\varrho_t(\chi)$ provides the characteristic function of $y_t$, denoted by $M(\chi,t)$. In fact, the characteristic function of a random variable is defined as the Fourier transform of its probability distribution. Hence, one has
\begin{equation}
\label{eq: chatacteristic function}
    M(\chi,t) \equiv E\left[ e^{i\chi y_t} \right] = \sum_y e^{i\chi y} P_t(y) = \tr[\varrho_t(\chi)]~,
\end{equation}
where we used Eq.~\eqref{eq: fourier transform of the memory-resolved state} and $P_t(y) = \tr[\varrho_t(y)]$. From the characteristic function, one can compute all the moments $E[y_t^j]\equiv \sum_y y^j P_t(y)$ of the stochastic memory for any $j=0,1,2,\cdots$ as 
\begin{equation}
    \label{eq: moments of the stochastic memory}
    E[y_t^j] = (-i\partial_\chi)^j \left.\tr[\varrho_t(\chi)] \right|_{\chi = 0}~.
\end{equation}

From the characteristic function \eqref{eq: chatacteristic function}, one can define the cumulant generating function as
\begin{equation}
    C(\chi,t)\equiv \log{M(\chi,t)} = \log{\left(\tr[\varrho_t(\chi)]\right)}~,
\end{equation}
and the cumulants $\langle\!\langle y_t^j\rangle\!\rangle \equiv (-i \partial_\chi)^j \left.C(\chi,t)\right|_{\chi = 0}$ become 
\begin{equation}
    \langle\!\langle y_t^j\rangle\!\rangle = (-i \partial_\chi)^j \left.\log{\left(\tr[\varrho_t(\chi)]\right)} \right|_{\chi = 0}~,
\end{equation}
where the first three cumulants are the mean, variance and skewness, respectively. Therefore, given the memory-resolved state $\varrho_t(y)$, one can compute its Fourier transform $\varrho_t(\chi)$, and one obtains all the moments and cumulants of the stochastic memory $y_t$.

\subsection{Memory correlations}
\label{subsec: memory correlations}

The next question we address concerns the correlations of the stochastic memory $y_t$ at different times. First, let us compute the conditional probabilities $P(y_{t'} = y' \mid y_t = y)$ of detecting a memory $y_{t'} = y'$ given a prior detection $y_t = y$ at time $t < t'$.

From Eq.~\eqref{eq: ensemble of the memory-conditioned states}, upon detecting $y_t = y$, the system's state at time $t$ becomes the memory-conditioned state $\rho_t(y) = \varrho_t(y)/\tr[\varrho_t(y)]$ [Eq.~\eqref{eq: memory-conditioned states}]. Moreover, given the information $y_t = y$, the probability distribution of the memory at time $t$ is $P_t(y' \mid y) = \delta_{y,y'}$, where $\delta_{a,b}$ denotes the Kronecker delta (or a Dirac delta if $y_t$ assume values in a continuous set).  
Therefore, the memory-resolved state at time $t$, conditioned on the previous event $y_t = y$, is given by
\begin{equation}
\label{eq: conditional memory-resolved state}
    \varrho_t(y' \mid y) \equiv \frac{\varrho_t(y)}{\tr[\varrho_t(y)]} \delta_{y,y'}~.
\end{equation}
Hence, the conditional probability $P(y_{t+\tau} = y' \mid y_t = y) \equiv P_\tau(y' \mid y)$ is given by
\begin{equation}
\label{eq: conditional probabilities}
    P_\tau(y' \mid  y) = \tr[\varrho_{t+\tau}(y' \mid y)]~,
\end{equation}
where $\varrho_{t+\tau}(y' \mid y)$ is the memory-resolved state evolved over a time interval $\tau \geq 0$ from the conditional state $\varrho_t(y' \mid y)$ according to the feedback dynamics [Eq.~\eqref{eq: deterministic eq 1}]. 

In particular, from Eq.~\eqref{eq: conditional probabilities}, one can compute the joint distribution $P(y_t = y, y_{t+\tau} = y')$ as
\begin{eqnarray}
    P(y_t = y, y_{t+\tau} = y') &=& P_t(y)\, P(y_{t+\tau} = y' \mid y_t = y) \nonumber\\
    &=& \tr[\varrho_t(y)]\,\tr[\varrho_{t+\tau}(y' \mid y)]~,
\end{eqnarray}
where we used $P_t(y) = \tr[\varrho_t(y)]$, and the conditional probability $P(y_{t+\tau} = y' \mid y_t = y) =\tr[\varrho_{t+\tau}(y' \mid y)]$.
From the conditional probabilities, one can introduce the two-point correlation function of the stochastic memory $y_t$ as
\begin{equation}
\label{eq: two-point correlation function of the memory}
    F(t,t+\tau) \equiv E[\delta y_t \delta y_{t+\tau} ] = E[y_t y_{t+\tau}] - \mu(t) \mu(t+\tau)~,
\end{equation}
where $\delta y_t \equiv y_t - \mu(t)$, and $\mu(t)=E[y_t]$ is the average value of the stochastic memory at time $t$. The term $E[y_t y_{t+\tau}]$ corresponds to the covariance function of $y_t$, and it is given by
\begin{eqnarray}
    E[y_t y_{t+\tau}] &=& \sum_{y,y'} y y' P_t(y_t=y,y_{t+\tau} = y')\\
    &=& \sum_{y,y'} y y' \tr[\varrho_t(y)]\,\tr[\varrho_{t+\tau}(y' \mid y)] ~.
\end{eqnarray}
Since $y_t$ is a classical stochastic process, we have $F(t, t+\tau) = F(t+\tau, t)$, and it is therefore sufficient to consider only $\tau \geq 0$. Note that $F(t, t) = \text{Var}[y_t]$ corresponds to the variance of $y_t$.

\section{No-feedback case and indirect feedback}
\label{sec: no feedback and indirect feedback}
All the formalism described above remains valid in the absence of feedback. Note that all quantities discussed depend solely on the memory-resolved state $\varrho_t(y)$, which is the solution of the feedback dynamics given by Eq.~\eqref{eq: deterministic eq 1}. In particular, the feedback can be removed by eliminating the memory dependence from the instruments, i.e., $\mathcal{M}_x(y) \rightarrow \mathcal{M}_x$. In this case, the evolution of the memory-resolved state from $t_n$ to $t_{n+1}$ becomes
\begin{equation}
    \label{eq: deterministic eq 2}
    \varrho_{n+1}(y) = \sum_{x',y'} \delta_{y, f_{n+1}(x',y')} \mathcal{M}_{x'} \varrho_n(y')~,
\end{equation}
and the full formalism developed in Sec.~\ref{sec:FCS_definitions} can be applied to analyze the statistics of the stochastic memory $y_t$.

Note that the statistical properties of the stochastic memory remain relevant even in the absence of feedback, as one can still solve Eq.~\eqref{eq: deterministic eq 2} to obtain the memory distribution $P_t(y)$. This distribution is experimentally accessible and, for instance, captures the statistics of detection outcomes under continuous monitoring limit $\delta t\rightarrow 0$. Therefore, such memory distribution may be used in metrology problems, where one aims to estimate a parameter from the observed stochastic data.

Now, we emphasize that Eq.~\eqref{eq: deterministic eq 1} can be straightforwardly generalized to the case of multiple causal memories. For instance, by considering two causal memories $y_n = f_n(x_n, y_{n-1})$ and $w_n = g_n(x_n, w_{n-1})$, the feedback dynamics becomes
\begin{eqnarray}
    \label{eq: deterministic eq 3}
    &\varrho_{n+1}(y,w) =\\ &\sum_{x',y',w'} \delta_{y, f_{n+1}(x',y')} \delta_{w, g_{n+1}(x',w')} \mathcal{M}_{x'}(y',w') \varrho_n(y',w')\nonumber~.
\end{eqnarray}
In particular, suppose that the feedback depends only on the memory $w_n$, but we are interested in the statistics of $y_n$. We refer to this situation as \textit{indirect feedback}.
In this case, the instruments depend only on $w$, and Eq.~\eqref{eq: deterministic eq 3} simplifies to
\begin{eqnarray}
    \label{eq: deterministic eq 4}
   & \varrho_{n+1}(y,w) =\\ &\sum_{x',y',w'} \delta_{y, f_{n+1}(x',y')} \delta_{w, g_{n+1}(x',w')} \mathcal{M}_{x'}(w') \varrho_n(y',w')\nonumber.
\end{eqnarray}

Finally, by solving Eq.~\eqref{eq: deterministic eq 4}, the statistics of the memory $y_n$ can be obtained by marginalizing the memory-resolved state $\varrho_n(y, w)$ over $w$. One has
\begin{equation}
    \varrho_n(y) = \sum_w \varrho_n(y, w)~,
\end{equation}
and $P(y_n = y) = \tr[\varrho_n(y)]$. Therefore, it illustrates how we can obtain the statistics of the memory $y_t$ under the condition that the feedback depends solely on a second memory $w_t$.

\section{Examples of Memory Functions}
\label{sec: examples of memory functions}
In this section, we present explicit examples of memory functions commonly used in feedback protocols. We consider two cases. 
The first involves a feedback protocol based on quantum-jump detections, where two distinct memories appear: one that records the last jump transition, and another that stores the time elapsed since that last jump. 
The second example corresponds to the \emph{current-resolved case}, in which the memory function equals the presently detected outcome, $y_n = x_n$ (or $y_t = x_t$ in the continuous limit).

\subsection{Jump-based feedback}
\label{sec: jump based feedback}
Let us consider a quantum system under continuous monitoring of its quantum jumps~\cite{Tutorial,Rosal2025DeterministicFeedback}, where detections are performed at regular times $t_n = n\,\delta t$, with $\delta t > 0$ an infinitesimal interval. 
In this scenario, the detected outcomes $x_n$ can be either $x_n = 0$ for a no-jump event, or $x_n = k$ when a jump $k \in \Sigma$ occurs at time $t_n$, where $\Sigma$ denotes the set of possible jump types.

The \emph{jump memory} $k_n$ is defined as
\begin{equation}
\label{eq: jump_filter}
k_n = x_n + k_{n-1}\delta_{x_n, 0} ~,
\end{equation}
with an initial condition $k_0 \in \Sigma$.
For instance, if a jump is detected at time $t_n$, so that $x_n = k$, the jump memory is updated to $k_n = x_n = k$. Conversely, if a no-jump event occurs, $x_n = 0$, the jump memory retains its previous value, $k_n = k_{n-1}$. In this way, $k_n$ always records the most recent jump transition.
Furthermore, the \emph{counting memory} is defined as
\begin{equation}
    \label{eq: counting_filter}
    \tau_n = \delta_{x_n,0}\,(\tau_{n-1} + \delta t)~.
\end{equation}
In this case, if a no-jump event is detected ($x_n = 0$), the counting signal is updated to $\tau_n = \tau_{n-1} + \delta t$, whereas when a jump is detected ($x_n = k$) it is reset to $\tau_n = 0$. 
Thus, $\tau_n$ keeps track of the time elapsed since the last jump transition.
In the continuous-monitoring limit $\delta t \to 0$, the jump and counting memories become continuous stochastic processes $k_t$ and $\tau_t$, representing, respectively, the last detected jump and the time elapsed since its occurrence.

A feedback protocol can be defined in terms of both the jump memory $k_t$ and the counting memory $\tau_t$. For example, depending on the last jump transition, one may modify the system Hamiltonian for a fixed period of time by turning on an external drive~\cite{Rosal2025DeterministicFeedback}. In this case, the system is governed by a stochastic Hamiltonian $H(k_t, \tau_t)$ that depends both on the last detected jump encoded in $k_t$ and on the time $\tau_t$ elapsed since its occurrence.
The feedback may also modify the jump operators that describe the system–environment interaction. For example, consider a qubit coupled to a thermal bath. The jump operators
\begin{equation}
\label{eq: jump operators thermal bath}
    L_{-} = \sqrt{\gamma(\bar{N}+1)}\,|g\rangle\langle e|, \qquad
L_{+} = \sqrt{\gamma \bar{N}}\,|e\rangle\langle g|,
\end{equation}
describe, respectively, the emission and absorption of a thermal photon from a bath at temperature $T$, where $\gamma>0$ is the coupling strength, $\bar{N} = (e^{\omega/T}-1)^{-1}$ is the Bose--Einstein distribution, and $\ket{g}$ and $\ket{e}$ are the ground and excited states separated by an energy gap $\omega$. 
By modifying the energy gap $\omega$ based on the last jump transition and on the time elapsed since its detection, one effectively alters the jump operators, resulting in $L_{-}(k_t,\tau_t)$ and $L_{+}(k_t,\tau_t)$.
This type of feedback mechanism can be experimentally implemented in quantum-dot setups~\cite{PhysRevLett.117.206803}.

Considering a general feedback protocol that modifies the system Hamiltonian $H(k_t,\tau_t)$ and/or the jump operators $\{L_q(k_t,\tau_t)\}_{q\in\Sigma}$, Ref.~\cite{Rosal2025DeterministicFeedback} showed that Eq.~\eqref{eq: deterministic eq 1}, together with the jump [Eq.~\eqref{eq: jump_filter}] and the counting [Eq.~\eqref{eq: counting_filter}] memories, reduces in the continuous-monitoring limit to the following deterministic equations for the memory-resolved state $\varrho_t(k,\tau)$:
\begin{eqnarray}
    \label{eq: deterministic equations combined signal1}
        \varrho_t(k,0) &=&  2\delta(t)\delta_{k,\bar{k}}\bar{\rho}_0+\sum_{q\in\Sigma}\int\limits_0^t d\tau \mathcal{J}_k(q,\tau) \varrho_t(q,\tau),\\
    \label{eq: deterministic equations combined signal2}
        \varrho_t(k,\tau) &=& G(k,\tau) \varrho_{t-\tau}(k,0),
    \end{eqnarray}
where $\mathcal{J}_k(q,\tau)\rho \equiv L_k(q,\tau) \rho L_k^\dagger(q,\tau)$ describes the jump channel $k\in\Sigma$, $\mathcal{L}_0(k,\tau)\equiv \mathcal{L}(k,\tau) - \sum_{q\in\Sigma}\mathcal{J}_q(k,\tau)$ is the no-jump Liouvillian,  $G(k,\tau)\equiv\mathcal{T}\left[e^{\int_0^\tau ds \mathcal{L}_0(k,s)}\right]$ is the propagator of the no-jump evolution, and $\mathcal{T}[\cdot]$ is the time-ordering operator.
The term $2\delta(t)\delta_{k,\bar{k}}\bar{\rho}_0$ describes the initial condition in which, at time $t=0$, the system is prepared in the state $\bar{\rho}_0$ and the counting memory is initialized in the state $\bar{k}$. We adopt the convention $\int_0^t d\tau~ \delta(t-\tau)=\frac{1}{2}$, which accounts for the Dirac delta function being located at the boundary of the integration interval.

Solving Eqs.~\eqref{eq: deterministic equations combined signal1} and \eqref{eq: deterministic equations combined signal2} for $\varrho_t(k,\tau)$, one obtains both the unconditional state $\bar{\rho}_t$ of the system and the probability distribution $P_t(k,\tau)$ of the jump and counting memories.
In this case, one has
\begin{eqnarray}
   \bar{\rho}_t &=& \sum_{k \in \Sigma} \int_0^t d\tau~ \varrho_t(k,\tau)~, \\
  P_t(k,\tau) &=& \tr[\varrho_t(k,\tau)]~,
\end{eqnarray}
where $P_t(k,\tau)$ gives the joint probability that the last jump transition was of type $k$ and that it occurred at time $t-\tau$. The distribution is normalized as $\sum_{k\in\Sigma} \int_0^t d\tau~P_t(k,\tau) = 1$ for any $t \geq 0$.

The feedback dynamics described by Eqs.~\eqref{eq: deterministic equations combined signal1} and \eqref{eq: deterministic equations combined signal2} cannot, in general, be expressed as a closed equation for the unconditional state.
However, this becomes possible when the feedback affects only the Hamiltonian, leaving the jump operators unchanged by the feedback action.
In this case, the unconditional state of the system evolves according to the following deterministic equation
~\cite{Rosal2025DeterministicFeedback}
\begin{equation}
    \label{eq: evolution of the unconditional state for Result 2}
    \bar{\rho}_t = G(\bar{k},t) \bar{\rho}_0+ \sum_{k\in\Sigma}\int_0^td\tau 
    G(k,\tau)
    \mathcal{J}_k \bar{\rho}_{t-\tau}~.
\end{equation}
For a no-jump Liouvillian whose eigenvalues have negative real parts, one has~\cite{Rosal2025DeterministicFeedback} $\lim_{t\to\infty} G(k,t)=0$. Hence, the steady state is obtained in the limit $t \to \infty$ and corresponds to the solution of the following algebraic equation
\begin{equation}
\label{eq: unconditional steady state for Result 2}
    \bar{\rho}_{\rm ss} = \Omega \bar{\rho}_{\rm ss},
    \qquad 
    \Omega = \sum_{k\in\Sigma}\int_0^\infty d\tau 
    G(k,\tau)
    \mathcal{J}_k.
\end{equation}
Finally, the steady-state memory-resolved state can be recovered from the unconditional steady state as
\begin{equation}
\label{eq: memory resolved state for jump based FB}
    \varrho_{\text{ss}}(k,\tau) = G(k,\tau)\mathcal{J}_k\bar{\rho}_{\text{ss}}.
\end{equation}
One may also consider a scenario in which the feedback depends solely on the last jump channel, irrespective of the time elapsed since its detection. This case is discussed in detail in Ref.~\cite{Rosal2025FCSwithFB}, where we further explore the connection between jump-based feedback and full counting statistics.

\subsection{Current-resolved case}
\label{sec: current resolved case}
\subsubsection{General instruments}
Let us consider the case in which the memory corresponds to the most recently detected outcome, $y_n = x_n$.
For simplicity, we assume that $x_n$ takes values in a discrete set; the continuous case can be treated analogously. This scenario is referred to as the \textit{current-resolved case}. For this choice of memory, the framework developed above yields the statistics of the outcomes at each step $t_n$. Since $y_n = f_n(x_n, y_{n-1})$ with $f(x, y) = x$ and an initial condition $y_0$, Eq.~\eqref{eq: deterministic eq 1} simplifies to
\begin{equation}
    \label{eq: deterministic equation for the current resolved state}
    \varrho_{n+1}(x) = \sum_{x'} \mathcal{M}_{x}(x') \varrho_n(x')~.
\end{equation}

The feedback steady state can be determined by reformulating Eq.~\eqref{eq: deterministic equation for the current resolved state} in matrix form. Note that this equation resembles the component form of a matrix equation. To make this structure explicit, let us define $\vec{\varrho}_n$ as a column vector whose components are $\varrho_n(x)$ for any possible outcome $x$, and introduce the two-index object $\Xi_{x,x'} \equiv  \mathcal{M}_x(x')$. 
Considering the standard matrix product, defined by $(A \cdot B)_{\alpha\beta} = \sum_z A_{\alpha z} B_{z\beta}$ for any double-indexed objects $X = [X_{\alpha\beta}]$, Eq.~\eqref{eq: deterministic equation for the current resolved state} becomes

\begin{equation}
    \vec{\varrho}_{n+1} = \Xi \cdot \vec{\varrho}_n~.
\end{equation}

Therefore, in the current-resolved case, the feedback steady state $\varrho_{\text{ss}}(x)$ then satisfies the algebraic equation $\vec{\varrho}_{\text{ss}} = \Xi \cdot \vec{\varrho}_{\text{ss}}$, which corresponds to an eigenvector-eigenvalue equation. 
By employing the vectorization technique~\cite{Tutorial}, $\Xi$ can be explicitly represented as a matrix and $\vec{\varrho}_n$ as a vector. The feedback steady state is thus obtained as the eigenvector of $\Xi$ associated with the eigenvalue 1.

For instance, Ref.~\cite{s9kj-lczm} implemented the following protocol: the position of a one-dimensional harmonic oscillator is continuously monitored, yielding outcomes $x_n$ at times $t_n = n\delta t$, with $\delta t$ much shorter than the characteristic timescales of the system. The detected position is then used to modify the external potential, thereby realizing a current-resolved feedback protocol. In their approach, stochastic sampling was used to describe this protocol, whereas Eq.~\eqref{eq: deterministic equation for the current resolved state} provides a deterministic equation whose solution directly yields the feedback dynamics.

\subsubsection{Feedback super-operators for a general measurement scheme}
Equation~\eqref{eq: deterministic equation for the current resolved state} describes a general feedback scheme based solely on the most recent detection event, applicable to any measurement protocol defined by the instruments $\mathcal{M}_x(x')$. We now particularize this equation to the case in which the feedback is implemented through a super-operator that depends explicitly on the latest detection outcome.

Let us consider that the measurements are performed at regular time intervals $t_n = n \, \delta t$, and are described by a set of Kraus operators $\{V_x\}_x$ satisfying the completeness relation $\sum_x V_x^\dagger V_x = \mathbb{I}$, where $\mathbb{I}$ is the identity. Between measurements, the system evolves from $t_n$ to $t_{n+1}$ under a dynamical map $e^{\delta t\mathcal{L}}$ with Liouvillian generator $\mathcal{L}$.
We assume that, conditional on the last detected outcome, a completely positive trace-preserving (CPTP) map $\mathcal{F}(x)$ -- referred to as the \textit{feedback super-operator} -- is applied to the system. The resulting instrument that describes this feedback scheme is then given by
\begin{equation}
\label{eq: instrument for current resolved with FB super-operator}
    \mathcal{M}_x \rho \equiv \mathcal{F}(x) M_x e^{\delta t\mathcal{L}} \rho~,
\end{equation}
where $M_x \rho \equiv V_x \rho V_x^\dagger$ represents the measurement channel associated with outcome $x$.

Therefore, the instrument in Eq.~\eqref{eq: instrument for current resolved with FB super-operator} describes the following sequence: first, the system evolves under the Liouvillian dynamics from $t_n$ to $t_{n+1}$; next, a measurement is performed at time $t_{n+1}$, yielding an outcome $x$; and finally, the feedback operation $\mathcal{F}(x)$ associated with this detected outcome is applied.
In Sec.~\ref{sec:application}, we illustrate this scheme with an example involving measurements on a qubit, where the feedback super-operator implements rotations in Hilbert space.
In this case, the deterministic equation \eqref{eq: deterministic equation for the current resolved state} becomes
\begin{eqnarray}
    \label{eq: deterministic equation for the current resolved state with FB super-operator}
    \varrho_{n+1}(x) &=&  \sum_{x'}\mathcal{F}(x) M_x e^{\delta t\mathcal{L}}\varrho_n(x')\\
    \label{eq: deterministic equation for the current resolved state with FB super-operator2}
    &=& \mathcal{F}(x) M_x e^{\delta t\mathcal{L}} \bar{\rho}_n~,
\end{eqnarray}
where $\sum_{x'} \varrho_{n}(x') = \bar{\rho}_n$.
Therefore, Eq.~\eqref{eq: deterministic equation for the current resolved state with FB super-operator2} yields
\begin{equation}
\label{eq: closed deterministic eq for current resolved case}
    \bar{\rho}_{n+1} = \sum_{x} \mathcal{F}(x) M_x e^{\delta t\mathcal{L}} \bar{\rho}_n~.
\end{equation}

Equation~\eqref{eq: closed deterministic eq for current resolved case} provides a closed evolution equation for the unconditional state of the system under a feedback protocol that depends solely on the currently detected outcome. However, as noted above, such a closed equation for the unconditional state is generally not available for arbitrary feedback dynamics. 
By solving Eq.~\eqref{eq: closed deterministic eq for current resolved case}, one obtains the unconditional state of the system at the discrete times $t_n$. In turn, Eq.~\eqref{eq: deterministic equation for the current resolved state with FB super-operator2} yields the memory-resolved state $\varrho_n(x)$, from which the outcome probability distribution is given by
\begin{equation}
\label{eq: probabilities current resolved}
    P(x_{n+1} = x) = \tr[\varrho_{n+1}(x)] = \tr[M_x e^{\delta t \mathcal{L}} \bar{\rho}_n]~.
\end{equation}

\subsubsection{Feedback super-operators for weak Gaussian measurements}

In this section, we focus on the case of a weak Gaussian measurement of a Hermitian operator $A$, which is described by the Kraus operators
\begin{equation}
\label{eq: kraus operators of gaussian measurements}
    V_x = \left(\frac{2\lambda \delta t}{\pi}\right)^{\frac{1}{4}}e^{-\lambda \delta t (x-A)^2}~,
\end{equation}
where $A$ is a given observable of the system, $\lambda$ denotes the measurement strength, and $\delta t > 0$ is an infinitesimal time interval~\cite{weakMeasurements1,weakMeasurements2}. 
The measurement outcomes are real numbers, $x \in \mathbb{R}$. In this continuous case, Eq.~\eqref{eq: closed deterministic eq for current resolved case} becomes
\begin{equation}
\label{eq: current resolved with WGM1}
    \bar{\rho}_{n+1} = \int_{-\infty}^\infty dx~ \mathcal{F}(x) M_x e^{\delta t\mathcal{L}} \bar{\rho}_n~.
\end{equation}

Let us consider the following feedback super-operator
\begin{equation}
\label{eq: feedback super operator for WGM}
    \mathcal{F}(z) = e^{ 2\sqrt{\lambda} z \delta t \mathcal{K} }~,
\end{equation}
where $\mathcal{K}\rho \equiv -i[F,\rho]$ represents a reversible transformation generated by a Hermitian operator $F$~\cite{WisemanBook}. 
Expanding $\mathcal{F}(x)$, $M_x$, and $e^{\delta t \mathcal{L}}$ to first order in $\delta t$, and taking the continuous-monitoring limit $\delta t \to 0$, Eq.~\eqref{eq: current resolved with WGM1} becomes (see the Supplemental Material of Ref.~\cite{Rosal2025DeterministicFeedback})
\begin{eqnarray}
\label{eq: current resolved with WGM2}
     \partial_t \bar{\rho}_t &=& \mathcal{L}\bar{\rho}_t+  \mathcal{D}\left[\sqrt{\lambda }A \right] \bar{\rho}_{t} \\
     &&+   \mathcal{D}[F]\bar{\rho}_{t}  -  i [F, \sqrt{\lambda}A\bar{\rho}_{t} + \bar{\rho}_{t} \sqrt{\lambda}A^\dagger]~\nonumber,
\end{eqnarray}
where $\mathcal{D}[L]\rho\equiv L\rho L^\dagger - 1/2 \{L^\dagger L,\rho\}$ denotes the Lindblad dissipator. 
The first term, $\mathcal{L}\bar{\rho}_t$, represents the intrinsic dynamical evolution of the system, while the term $\mathcal{D}\!\left[\sqrt{\lambda}\,A\right]$ arises from the continuous monitoring of $A$ via the weak Gaussian measurement. The remaining two terms originate from the continuous feedback action: a dissipative component, $\mathcal{D}[F]$, and a coherent contribution $-i[F,\cdot\,]$. 
Equation~\eqref{eq: current resolved with WGM2} was originally developed in Ref.~\cite{WisemanMilburn1993_Homodyne} in the context of diffusive homodyne detection. 
Ref.~\cite{PatrickQFPME} extended this equation to diffusive measurements induced by weak Gaussian measurements, and more recently Ref.~\cite{Rosal2025DeterministicFeedback} showed that this feedback master equation can be derived directly from Eq.~\eqref{eq: deterministic eq 1}.

\section{Applications}
\label{sec:application}

In the next two examples, we consider a qubit (two-level system) with Hamiltonian $H$, defined on a Hilbert space with orthonormal basis $\{\ket{g}, \ket{e}\}$, where $\ket{g}$ and $\ket{e}$ denote the ground and excited states, respectively, with energy gap $\omega$.  
We assume the qubit is coupled to a thermal bath at temperature $T$, characterized by a Bose-Einstein distribution $\bar{N} = (e^{\omega_{}  /  T} - 1)^{-1}$.
The interaction between the system and the thermal bath, in the weak-coupling limit, is described by the Lindblad master equation $\partial_t \rho(t) = \mathcal{L} \rho(t)$, where the Liouvillian generator $\mathcal{L}$ is given by 
\begin{equation}
\label{eq: thermal dynamics}
\mathcal{L} \rho(t) = -i[H, \rho(t)] + \mathcal{D}[L_+]\rho(t) + \mathcal{D}[L_-]\rho(t)~.
\end{equation}
The jump operators $L_{\pm}$ are defined in Eq.~\eqref{eq: jump operators thermal bath}, and they describe the system’s ability to absorb or emit photons from the environment.
Note that both the Hamiltonian $H$ and the coupling strength $\gamma$ in Eq.~\eqref{eq: thermal dynamics} are expressed in units of frequency ($\hbar = 1$).
The jump channels are defined as $\mathcal{J}_{\pm1}\rho \equiv L_\pm \rho L_\pm^\dagger$, and the no-jump Liouvillian is given by $\mathcal{L}_0 = \mathcal{L} - (\mathcal{J}_{-1} + \mathcal{J}_{+1})$.

Considering only the thermal coupling described by Eq.~\eqref{eq: thermal dynamics}, the system typically evolves toward a steady state in which the population of the ground state $\ket{g}$ exceeds that of the excited state $\ket{e}$, $P_e < P_g$. Furthermore, if a Rabi drive is present in the system Hamiltonian, the interaction with the thermal bath suppresses the Rabi oscillations in the steady state.
In Sec.~\ref{subsec: inversion protocol}, we present a jump-based feedback protocol (originally developed in Ref.~\cite{Rosal2025DeterministicFeedback}) that, based on jump detections, achieves a steady state with $P_e > P_g$ and is referred to as the \emph{inversion protocol}.
In the second example, Sec.~\ref{subsec: Rabi stabilization} presents a discrete feedback protocol based on projective measurements (implemented experimentally in Ref.~\cite{DiscreteFeedback3}) that stabilizes Rabi oscillations against the thermal effects of the environment.
We then apply the tools developed in Sec.~\ref{subsec: Hybrid states} and \ref{sec:FCS_definitions} to compute, for example, the memory probability distribution, the mutual information between the memory and the system, the covariance between the memory and the system's observables, and the average values of the memory function.

\subsection{Inversion protocol with quantum jump detections}
\label{subsec: inversion protocol}
\begin{figure}
    \centering
     \includegraphics[scale=0.35]{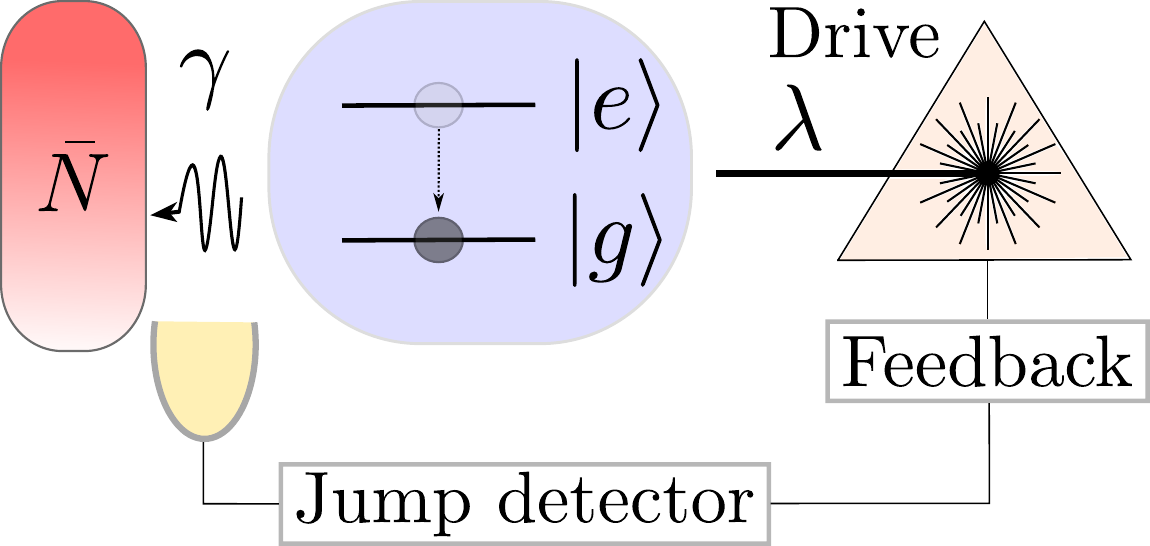}
    \caption{ Schematics of the inversion protocol. After detecting a decay event $\ket{e}\!\to\!\ket{g}$, the feedback-activated drive acts to steer the system back toward the excited state $\ket{e}$.
 }
    \label{fig: diagram inversion protocol}
\end{figure}

Considering the rotating frame at frequency $\omega_d$, the qubit Hamiltonian reads $  H = -\frac{\Delta}{2}\,\sigma_z$, where $\sigma_{x,y,z}$ are the Pauli matrices and $\Delta\equiv\omega-\omega_d$. In our convention the ground state $\ket{g}$ has energy $-\Delta/2$ and the excited state $\ket{e}$ has energy $+\Delta/2$, with $\sigma_z = \ket{g}\bra{g} - \ket{e}\bra{e}$.
A thermal emission corresponds to the incoherent jump $\ket{e}\to\ket{g}$ and can be detected, for instance, through photon detectors~\cite{PhysRevLett.57.1699,PhysRevLett.57.1696,PhysRevLett.56.2797}. 
Supposing the continuous monitoring of these quantum jumps, we define the feedback protocol as follows. 
If we detect a thermal emission ($\ket{e}\!\to\!\ket{g}$) and observe no subsequent detection for a time $\tau_0$, 
an external drive is turned on for an interval $\tau_1$, and the Hamiltonian becomes $H = -\frac{\Delta}{2}\sigma_z + \lambda\sigma_x$.
In summary, once an emission occurs and a time $\tau_0$ has elapsed without subsequent jump detections, the drive is activated and induces Rabi oscillations for a duration $\tau_1$, transferring population from $\ket{g}$ back to $\ket{e}$ (see Fig.~\ref{fig: diagram inversion protocol}).

In this setup, two types of jumps may occur: \(k=-1\) for an emission event \(\ket{e}\!\to\!\ket{g}\), and \(k=+1\) for an absorption event \(\ket{g}\!\to\!\ket{e}\).
Accordingly, the feedback protocol leads to a stochastic Hamiltonian that depends on both the jump memory and the counting memory, defined in Eqs.~\eqref{eq: jump_filter} and \eqref{eq: counting_filter}, and is given by
\begin{equation}
\label{eq: stochastic hamiltonian for inversion prot}
    H(k_t,\tau_t) = -\frac{\Delta}{2} \sigma_z + \lambda~\delta_{-1,k_t}\theta(\tau_t - \tau_0) \theta(\tau_0+\tau_1 - \tau_t)~ \sigma_x~,
\end{equation}
where $\delta_{a,b}$ is the Kronecker delta, and $\theta(x)$ is the Heaviside function. 
Here, $H(k_t,\tau_t)$ specifies the activation of the external drive whenever the last jump channel is an emission ($k_t = -1$) and the time elapsed since its detection satisfies $\tau_0<\tau_t<\tau_0+\tau_1$.

\begin{figure}
    \centering
     \includegraphics[scale=0.4]{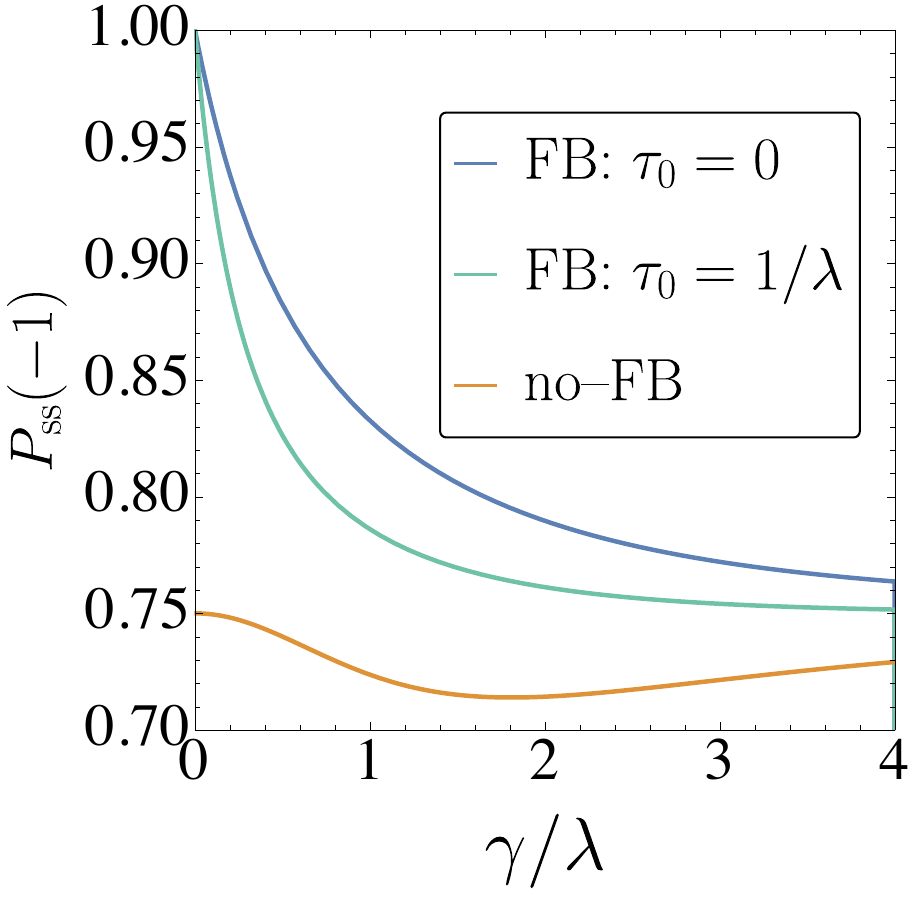}
    \caption{ Probability distribution $P_{\text{ss}}(k)$ of the jump memory $k_t$ in the steady state. 
    The jump memory $k_t$ can take the values $-1$ (decay $\ket{e}\to\ket{g}$) and $+1$ (excitation $\ket{g}\to\ket{e}$), so that $P_{\text{ss}}(-1)$ is the probability that the last jump in the steady state corresponds to a decay, and $P_{\text{ss}}(+1) = 1 - P_{\text{ss}}(-1)$. 
    The no-feedback case corresponds to an external drive that is always on, with dynamics described by the Liouvillian in Eq.~\eqref{eq: thermal dynamics}.
    We consider $\tau_1 = \tau_1^\text{opt}$ [Eq.~\eqref{eq: optimal drive time}], and $\bar{N} = 0.5$. 
    }
    \label{fig: jump memory distribution}
\end{figure}

The steady-state solution $\bar{\rho}_{\mathrm{ss}}$ of this feedback protocol was fully characterized in Ref.~\cite{Rosal2025DeterministicFeedback}, where it was obtained as the solution of Eq.~\eqref{eq: unconditional steady state for Result 2}. 
For instance, Ref.~\cite{Rosal2025DeterministicFeedback} showed that the optimal drive duration $\tau_1^\text{opt}$, which maximizes the probability of the excited state in the stationary regime, is given by
\begin{equation}
\label{eq: optimal drive time}
    \tau_1^{\text{opt}} = \frac{2\left[2\pi +\arctan{\left( \frac{p\sqrt{16-p^2}}{8-p^2} \right)}-\pi\theta(\sqrt{8}-p)\right]}{\lambda\sqrt{16-p^2}} ~,
\end{equation}
where $p = \gamma/\lambda$ quantifies the competition between the drive strength $\lambda$ and the thermal coupling $\gamma$ in the feedback dynamics.

In what follows, we concentrate on the properties of the jump memory $k_t$ that underlies the feedback dynamics, and we consider the optimal drive duration $\tau_1 = \tau_1^{\text{opt}}$.
The steady-state distribution resolved over both the jump and counting memories follows from Eq.~\eqref{eq: memory resolved state for jump based FB}. 
In particular, the joint steady-state probability density of observing a jump $k$ followed by no-jump detections for a time interval $\tau$ is given by
\begin{equation}
    P_{\text{ss}}(k,\tau)
    = \tr[\varrho_{\text{ss}}(k,\tau)]
    = \tr\!\big[\,G(k,\tau)\,\mathcal{J}_k\,\bar{\rho}_{\text{ss}}\,\big].
\end{equation}
To obtain the jump-memory–resolved state, one marginalizes over $\tau$, giving
\begin{eqnarray}
    \label{eq: jump-memory state 1}
    \varrho_{\text{ss}}(k) &=& \int_0^\infty \nonumber
    d\tau~\varrho_{\text{ss}}(k,\tau)\\ &=& \left(\int_0^\infty d\tau~ G(k,\tau)\right)\mathcal{J}_k\bar{\rho}_{\text{ss}}~,
\end{eqnarray}
where $G(+1,\tau) = e^{\tau \mathcal{L}_0^{\text{off}}}$, with $\mathcal{L}_0^{\text{off}}$ being the no-jump Liouvillian associated with the Hamiltonian $H_\text{off} \equiv -\frac{\Delta}{2}\, \sigma_z$, and
\begin{equation}
\label{eq: no-jump propagator}
    G(-1,\tau) = \begin{cases}
        e^{\tau\mathcal{L}_0^{\text{off}}}~&\text{if }\tau\leq\tau_0,\\
        e^{(\tau-\tau_0)\mathcal{L}_0^{\text{on}}}e^{\tau_0\mathcal{L}_0^{\text{off}}}~&\text{if }\tau\in [\tau_0,\tau_0+\tau_1],\\
        e^{(\tau-\tau_0-\tau_1)\mathcal{L}_0^{\text{off}}} e^{\tau_1\mathcal{L}_0^{\text{on}}}e^{\tau_0\mathcal{L}_0^{\text{off}}}~&\text{if }\tau_0+\tau_1\leq\tau,
                \end{cases}
\end{equation}
where $\mathcal{L}_0^{\text{on}}$ is the no-jump Liouvillian associated with the Hamiltonian $H_\text{on} \equiv -\frac{\Delta}{2}\sigma_z + \lambda \sigma_x$.
From this solution of the no-jump propagator $G(k,\tau)$, one can evaluate the integral in Eq.~\eqref{eq: jump-memory state 1}, yielding $\varrho_{\text{ss}}(\pm 1) = \Lambda_{\pm} \bar{\rho}_{\text{ss}}$, where
\begin{eqnarray}
    \Lambda_{+} &\equiv& -[\mathcal{L}_0^{\text{off}}]^{-1}\mathcal{J}_{1},\\
     \Lambda_{-} &\equiv & \Big\{ -\left[\mathcal{L}_0^\text{off}\right]^{-1} + \left(\left[\mathcal{L}_0^\text{off}\right]^{-1} -\left[\mathcal{L}_0^\text{on}\right]^{-1}\right) e^{\tau_0\mathcal{L}_0^\text{off}}  \nonumber\\&&+ \left(\left[\mathcal{L}_0^\text{on}\right]^{-1} -\left[\mathcal{L}_0^\text{off}\right]^{-1}\right) e^{\tau_1\mathcal{L}_0^\text{on}}  e^{\tau_0\mathcal{L}_0^\text{off}}    \Big\}\mathcal{J}_{-1},
\end{eqnarray}
where $[A]^{-1}$ denotes the inverse of an operator $A$. Hence, from the jump-memory–resolved state $\varrho_{\text{ss}}(k)$, one can apply all the tools introduced in Sec.~\ref{subsec: Hybrid states}.

\begin{figure}
    \centering
     \includegraphics[scale=0.4]{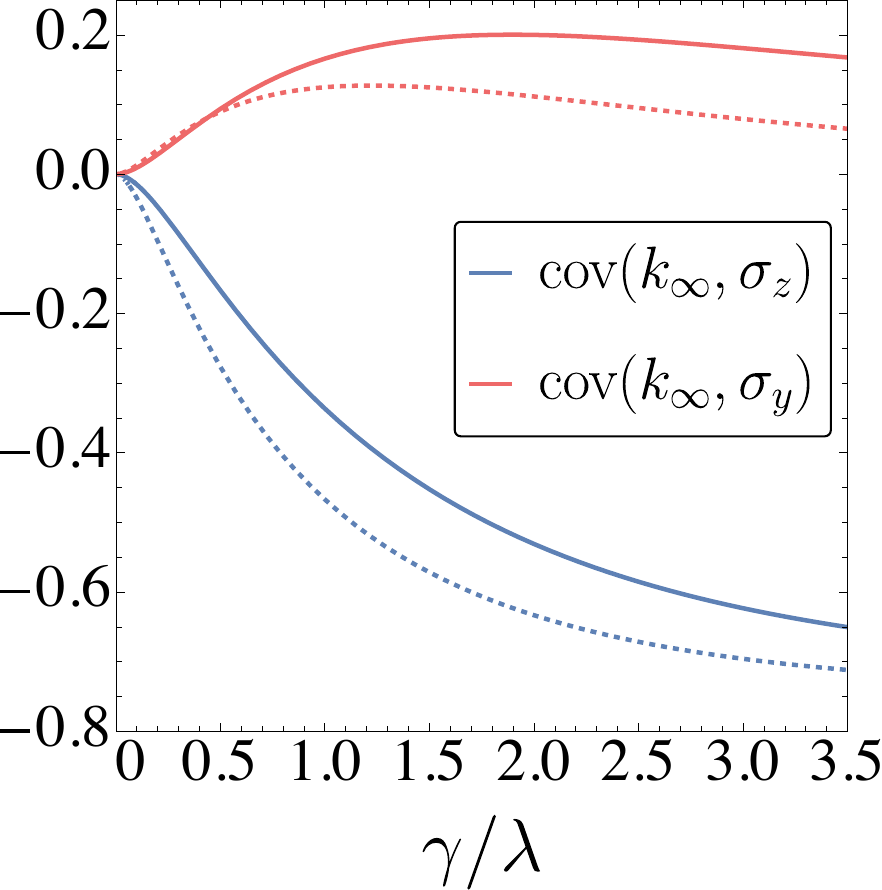}
    \caption{ Covariance between the Pauli operators $\sigma_z$, $\sigma_y$ and the jump memory $k_t$ in the steady state, as defined in Eq.~\eqref{eq: def memory covariance}. 
    The blue lines correspond to the covariance between $\sigma_z$ and $k_t$, whereas the red lines correspond to the covariance between $\sigma_y$ and the jump memory. 
    The parameters used are $\bar{N} = 0.5$, with $\tau_0 = 0$ for solid lines and $\tau_0 = 1/(2\lambda)$ for dashed lines, and $\tau_1 = \tau_1^\text{opt}$ [Eq.~\eqref{eq: optimal drive time}].
    }
    \label{fig: jump memory covariance}
\end{figure}

The jump-memory probability distribution in the steady state is given by $P_{\text{ss}}(\pm 1) = \tr[\varrho_{\text{ss}}(\pm 1)]$. 
$P_{\text{ss}}(k)$ represents the probability that the last jump channel was $k$, with $k = -1$ corresponding to an emission ($\ket{e}\to\ket{g}$) and $k = +1$ corresponding to an absorption ($\ket{g}\to\ket{e}$) of a photon from the environment to the qubit. 
In Fig.~\ref{fig: jump memory distribution}, we show the probability $P_{\text{ss}}(-1)$, with $P_{\text{ss}}(+1) = 1 - P_{\text{ss}}(-1)$.
Note that $P_{\text{ss}}(-1)$ approaches one in the strong-drive limit $\lambda \gg \gamma$. In this regime, the feedback action keeps the system predominantly in the excited state, so the only possible transition corresponds to a decay $\ket{e}\to\ket{g}$, represented by $k = -1$.
As the time delay $\tau_0$ increases, the performance of the inversion protocol decreases, and one observes that the probability $P_{\text{ss}}(-1)$ decreases as a function of $\gamma/\lambda$. 
This occurs because a larger $\tau_0$ allows the system to spend more time in the ground state $\ket{g}$, making it more likely to detect absorptions, which deteriorates the effectiveness of the protocol.

Figure~\ref{fig: jump memory covariance} shows the covariance [Eq.~\eqref{eq: def memory covariance}] between the Pauli operators $\sigma_z = \ket{g}\bra{g} - \ket{e}\bra{e}$ and $\sigma_y$, and the jump memory $k_t$ in the steady state. In our convention, the ground state $\ket{g}$ is the eigenstate of $\sigma_z$ with eigenvalue $+1$ [see Eq.~\eqref{eq: stochastic hamiltonian for inversion prot}].
Note that the covariance between $\sigma_z$ and $k_t$ is negative: when $k_t$ is above its average, $\sigma_z$ tends to be below its average. 
For example, as the number of emissions ($k_t = -1$) increases, $\sigma_z$ tends to move toward the ground state $\ket{g}$, which has eigenvalue $+1$.
The solid lines correspond to $\tau_0 = 0$, and the dashed lines represent $\tau_0 = 1/(2\lambda)$. 
As $\tau_0$ decreases, the protocol becomes more efficient: when an emission ($k_t = -1$) is detected, the feedback mechanism drives the system back to the excited state $\ket{e}$, which is the eigenstate of $\sigma_z$ with eigenvalue $-1$.
Hence, the feedback increases the covariance between $k_t$ and $\sigma_z$. 
By contrast, the opposite behavior occurs between $\sigma_y$ and $k_t$, due to the feedback action implemented by the drive $\sigma_x$ in Eq.~\eqref{eq: stochastic hamiltonian for inversion prot}.

\begin{figure}
    \centering
     \includegraphics[scale=0.4]{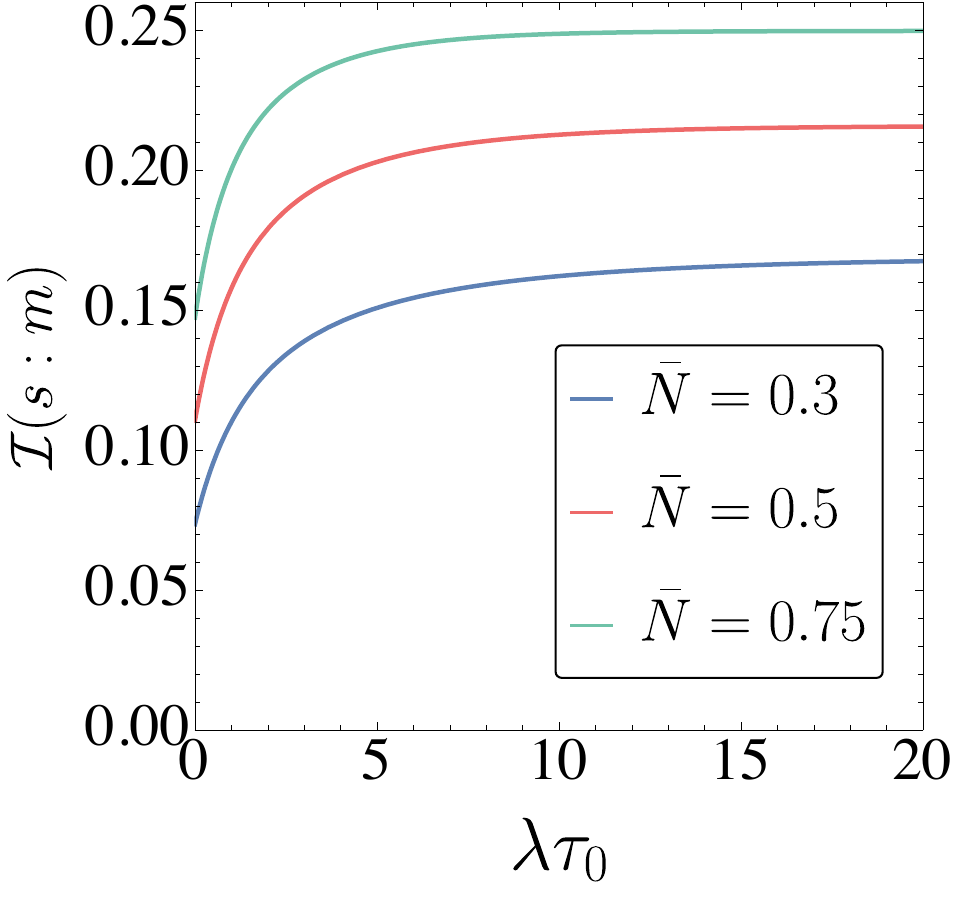}
    \caption{  Mutual information between the system and the jump memory in the steady state as a function of $\lambda\tau_0$, for $\gamma/\lambda = 0.5$ and different bath temperatures. }
    \label{fig: jump memory mutual information}
\end{figure}

Figure~\ref{fig: jump memory mutual information} shows the mutual information between the system and the jump memory $k_t$, as defined in Eq.~\eqref{eq: mutual classical-quantum information}. 
In this case, the mutual information quantifies the uncertainty about the system's state given the observation of a jump.
Note that this feedback protocol drives the system toward the excited state, thereby reducing the uncertainty of the system's state given that a jump was detected. 
Consequently, the more efficiently the feedback stabilizes the qubit in the excited state, the smaller the resulting uncertainty.
The efficiency of the feedback can be enhanced, for instance (see Fig.~3(b) and (c) of \cite{Rosal2025DeterministicFeedback}), by decreasing the delay time $\tau_0$, increasing the drive strength $\lambda$ relative to the thermal coupling $\gamma$, or reducing the Bose-Einstein occupation $\bar{N}$ of the bath.
In addition, we note that the mutual information increases with temperature. This occurs because, at higher temperatures, jump events become more frequent. Consequently, when examining the memory at a given time, the last jump channel is likely to have occurred more recently and therefore carries more information about the state than when the last jump lies far in the past.

\subsection{Stabilizing Rabi oscillations of a dissipative qubit with projective based feedback}
\label{subsec: Rabi stabilization}

\begin{figure}
    \centering
     \includegraphics[scale=0.25]{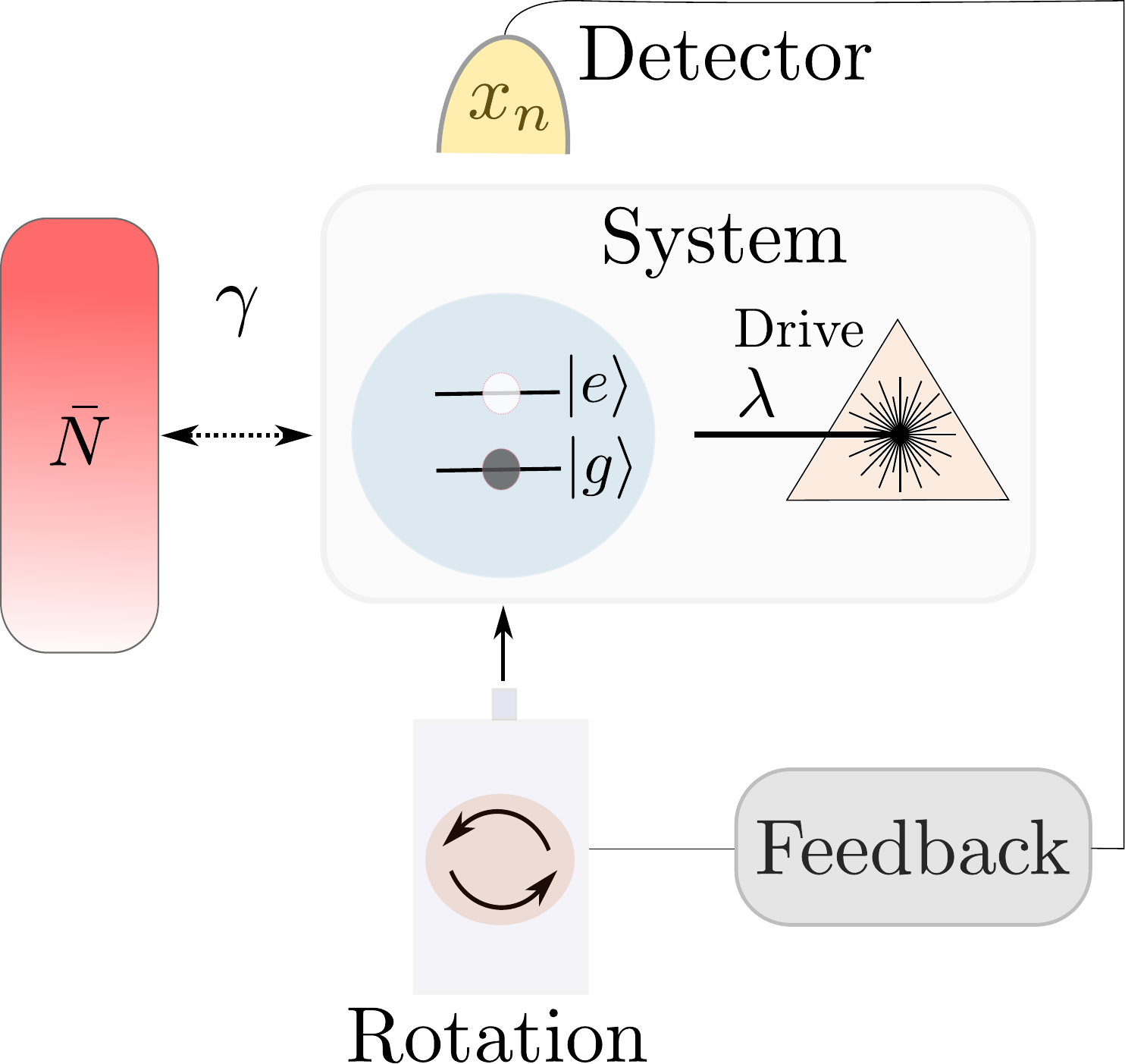}
    \caption{Diagram of the stabilization of Rabi oscillations. A qubit is coupled to both a thermal bath and a coherent external drive, and is subjected to projective measurements of its states $\ket{e}$ and $\ket{g}$ at regular intervals $\delta t$.
    Whenever the qubit is found in the excited state $\ket{e}$, a $\pi$-rotation around the $x$-axis of the Bloch sphere is applied, bringing the system to the ground state $\ket{g}$.
    }
    \label{fig: Rabi stabilization diagram}
\end{figure}

In this application, we focus on the stabilization of Rabi oscillations in a two-level system coupled to a thermal bath via discrete feedback (see Fig.~\ref{fig: Rabi stabilization diagram}), as experimentally demonstrated in Ref.~\cite{DiscreteFeedback3}.
We consider projective measurements of the states $\ket{e}$ and $\ket{g}$, such that the outcomes at time $t_n = n~\delta t$ are $x_n = +1$ (excited state $\ket{e}$) or $x_n = -1$ (ground state $\ket{g}$), and $n = 1,2,\cdots$.
The associated projectors are $\Pi_+ \equiv |e\rangle\langle e|$ and $\Pi_- \equiv |g\rangle\langle g|$, respectively, and the measurement process is described by the measurement channels $M_\pm \rho = \Pi_\pm \rho \Pi_\pm^\dagger$.

We focus on the current-resolved case, where the memory corresponds to the most recent detection outcome, $y_n = x_n$. The feedback is defined as follows: if the system is detected in the excited state ($x_n = +1$), a $\pi$-rotation is instantaneously applied to return the system to the ground state; if $x_n = -1$, no operation is applied. Accordingly, the feedback super-operator takes the form
\begin{equation}
\label{eq: FB super-operator with rotations}
    \mathcal{F}(x) =  \begin{cases}
        I_d, & \text{if } x = -1~,\\
        R(\pi), & \text{if } x = +1~,
    \end{cases}
\end{equation}
where $R(\pi)\rho = e^{-i\pi\sigma_x/2} \rho e^{i\pi\sigma_x/2}$ is the unitary channel implementing a rotation of the state $\rho$ by angle $\pi$ around the $x$-axis of the Bloch sphere, and $I_d$ denotes the identity super-operator, $I_d(\rho) = \rho$. 
This strategy corresponds to an ideal scenario in which there is no feedback delay. However, a finite time delay between the detection and the feedback action can also be incorporated~\cite{Rosal2025DeterministicFeedback}.

This feedback strategy is based solely on the presently detected outcome and is therefore fully described by Eq.~\eqref{eq: closed deterministic eq for current resolved case}. 
Using the projective measurements $M_\pm$, the thermal coupling described by the Liouvillian in Eq.~\eqref{eq: thermal dynamics}, and the feedback super-operator defined in Eq.~\eqref{eq: FB super-operator with rotations}, one obtains $\bar{\rho}_{n+1} = \ket{g}\bra{g}$ for $n = 0,1,\cdots$. 
Hence, the system is reset to the ground state after the measurement process, regardless of the detected outcome. 
Consequently, the dynamics in each interval $\delta t$ becomes independent of the system's past history.
This corresponds to an example in which the feedback transforms the system's dynamics into a renewal process, where the evolution from $t_n$ to $t_{n+1}$ is independent of the previous states, as the system always starts each interval $t_n$ in the ground state $\ket{g}$.

\begin{figure}
    \centering
     \includegraphics[scale=0.4]{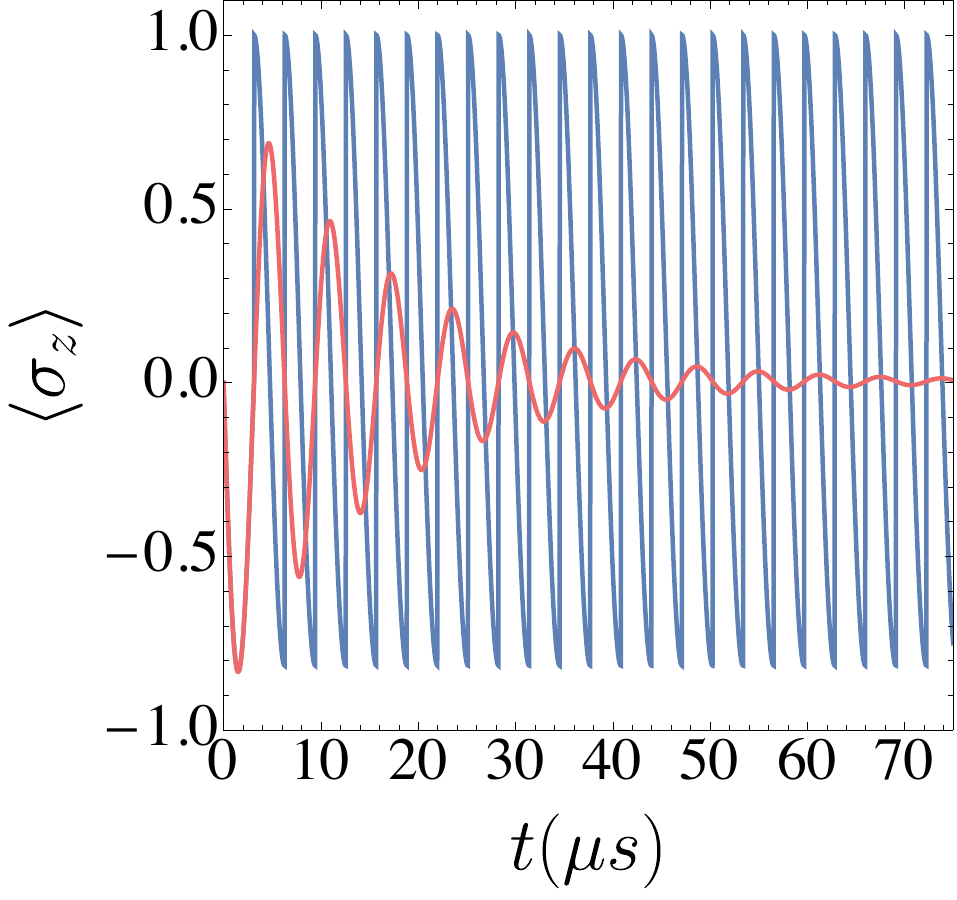}
    \caption{  Dynamical behavior of the qubit populations under the stabilization of Rabi oscillations.
    The red line corresponds to the no-feedback case, where the qubit is coupled to a thermal bath and driven by a coherent external field, without measurements, and the evolution is governed by Eq.~\eqref{eq: thermal dynamics} with the Hamiltonian given in Eq.~\eqref{eq: rabi H}. 
    We adopt parameters similar to those used in the experimental implementation of Ref.~\cite{DiscreteFeedback3}, with qubit frequency $\omega_{}/2\pi = 3.57$ GHz, bath temperature $T = 46$ mK, and coupling strength $\gamma = 80$ kHz. 
    We consider a coherent external drive with zero detuning, i.e., $\Delta = 0$ in Eq.~\eqref{eq: rabi H}, and a drive strength $\lambda = 500$ kHz. The system is initially prepared in the pure state $(\ket{e} + \ket{g})/\sqrt{2}$, and the time between measurements is set to $\delta t = \pi/(2\lambda)$. 
    }
    \label{fig: population Rabi stabilization}
\end{figure}

Let us consider that the qubit is driven by a continuous external field with frequency $\omega_d$. By applying the rotating wave approximation and moving to a rotating frame at frequency $\omega_d$, the qubit Hamiltonian becomes
\begin{equation}
\label{eq: rabi H}
    H = -\frac{\Delta}{2} \sigma_z + \lambda \sigma_x~,
\end{equation}
where $\Delta \equiv \omega_{} - \omega_d$, and $\lambda$ denotes the drive strength. 
In the present case, we assume a resonant drive, $\omega_d = \omega_{}$, so that $\Delta = 0$. Additionally, we consider that the system is initially prepared in the pure state $(\ket{e} + \ket{g}) / 2$.
In Fig.~\ref{fig: population Rabi stabilization}, we show the dynamical evolution of the average $\braket{\sigma_z}$, obtained using parameters similar to those in the experimental implementation of Ref.~\cite{DiscreteFeedback3}. 
In the absence of measurements, the thermal coupling suppresses the Rabi oscillations (red curve), and the system evolves toward the maximally mixed state, where $\braket{\sigma_z} = 0$.
In contrast, this feedback protocol effectively preserves the Rabi oscillations between $\ket{g}$ and $\ket{e}$ against thermal decay.

The coherent Rabi oscillation, corresponding to a two-level system without any thermal bath and with Hamiltonian given by Eq.~\eqref{eq: rabi H}, exhibits persistent oscillations of $\braket{\sigma_z}$ between $-1$ and $1$. 
On the other hand, the feedback-stabilized dynamics shown in Fig.~\ref{fig: population Rabi stabilization} does not achieve $\braket{\sigma_z} = -1$. The reason is that, although the oscillations are stabilized, dissipation induced by the thermal bath still occurs during each time interval $\delta t$ of the system evolution.

\begin{figure}
    \centering
     \includegraphics[scale=0.4]{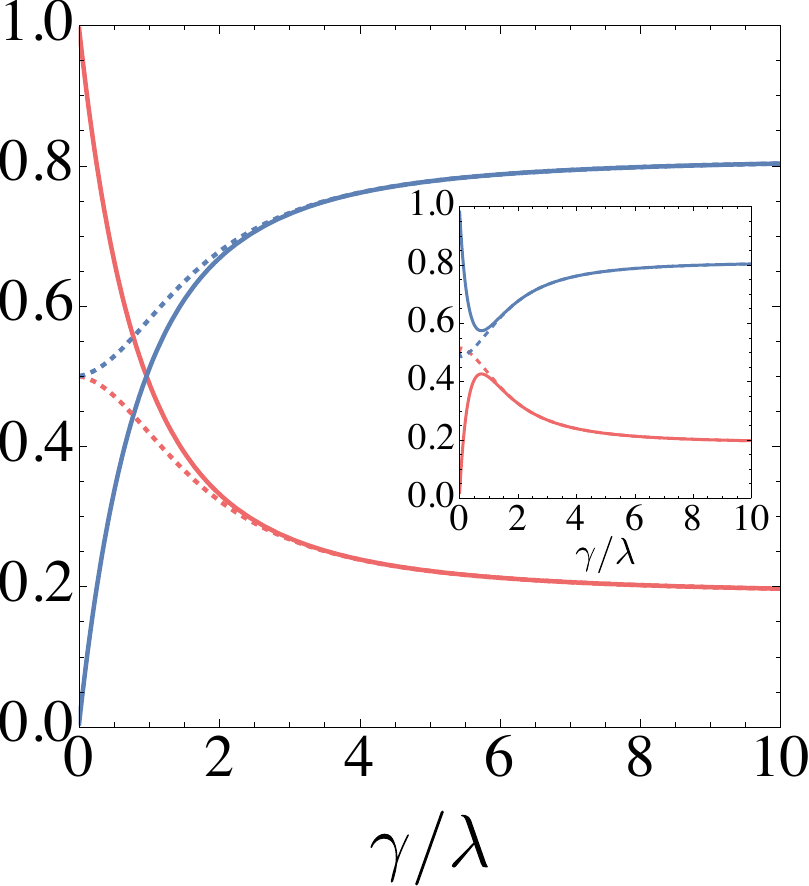}
    \caption{
    Probability distributions $P(x_n = +1) = \bra{e}\, e^{\delta t \mathcal{L}}\!\left( \ket{g}\!\bra{g} \right) \ket{e}$ (red curves) and $P(x_n = -1) = \bra{g}\, e^{\delta t \mathcal{L}}\!\left( \ket{g}\!\bra{g} \right) \ket{g}$ (blue curves) for $n \geq 2$.
    The parameters used were $\bar{N} = 0.3$, $\lambda \delta t = 1.5$ for the main panel and $\lambda \delta t = 3$ for the inset. Dashed lines correspond to the no-feedback case, where only the measurements are performed and no rotation is applied.}
    \label{fig: outcomes prob dist Rabi}
\end{figure}

From Eq.~\eqref{eq: probabilities current resolved}, the probability distribution at each step for this projective measurement scheme corresponds to the populations of the dynamical state $e^{\delta t \mathcal{L}} \bar{\rho}_n$.
Hence, for the first measurement, one has $P(x_1 = +1) = \bra{e} e^{\delta t \mathcal{L}} \rho_0 \ket{e}$ and $P(x_1 = -1) = \bra{g} e^{\delta t \mathcal{L}} \rho_0 \ket{g}$, where $\rho_0$ is the initial state of the system.
Once $\bar{\rho}_n = \ket{g}\bra{g}$ for any $n =1,2,\cdots$, one has 
$P(x_n = +1) = \bra{e} e^{\delta t \mathcal{L}} \left( \ket{g}\bra{g} \right) \ket{e}$ and 
$P(x_n = -1) = \bra{g} e^{\delta t \mathcal{L}} \left( \ket{g}\bra{g} \right) \ket{g}$ for $n \geq 2$.
Figure~\ref{fig: outcomes prob dist Rabi} shows $P(x_n)$ for $n \geq 2$ as a function of $\gamma/\lambda$, for different values of the measurement interval $\delta t$.
Note that the drive strength and measurement interval $\delta t$ can be adjusted such that, in the strong drive regime $\lambda \gg \gamma$, the feedback ensures that $x_n = +1$ or $x_n = -1$ is detected with high certainty.
The dashed lines correspond to the same probabilities in the no-feedback case, where the system is measured at each time step $t_n$, but no rotation is applied.

\begin{figure}
    \centering
     \includegraphics[scale=0.4]{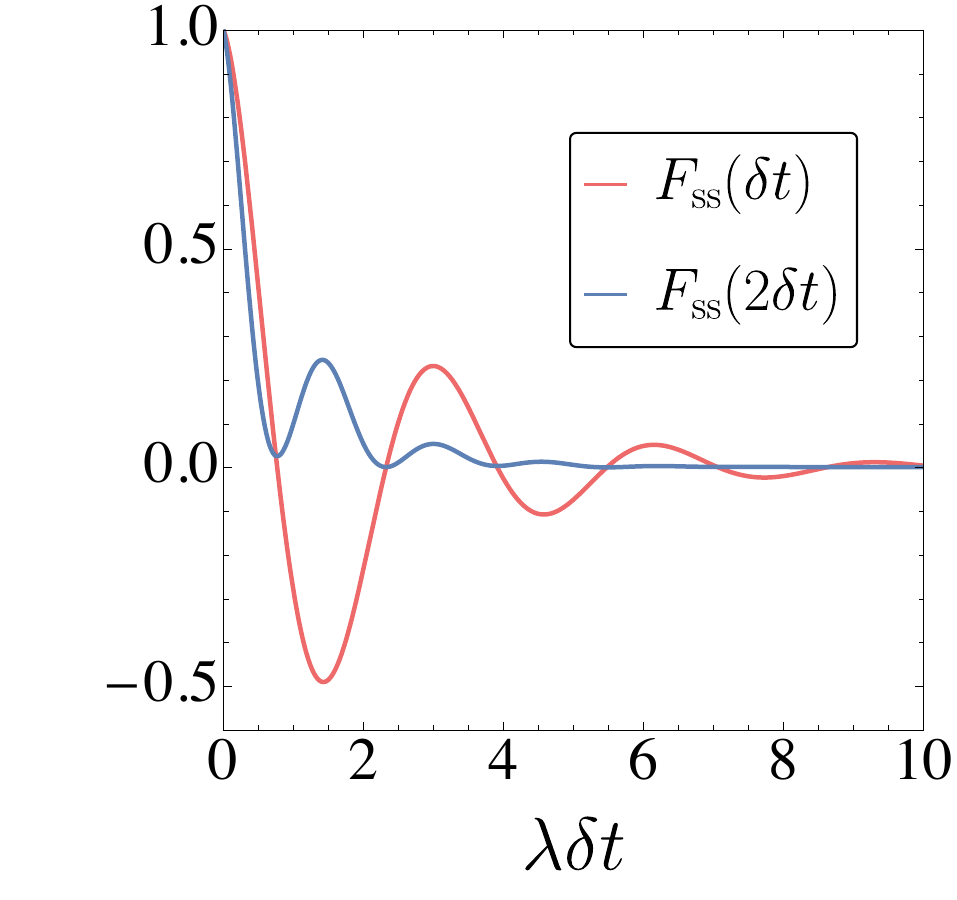} 
  \caption{
    Two-point correlation function in the steady state for the no-feedback case, for two measurements separated by an interval $\delta t$ and $2\delta t$, denoted by $F_{\text{ss}}(\delta t)$ and $F_{\text{ss}}(2\delta t)$, respectively. The parameters used were $\bar{N} = 0.3$ and $\gamma/\lambda = 0.4$.
}
    \label{fig: corr function no-FB Rabi}
\end{figure}

Once the feedback resets the system to the ground state at each time step $t_n$, the evolutions over different intervals become uncorrelated, and the measurements at times $t_n$ are independent and identically distributed.
In particular, the two-point correlation functions [Eq.~\eqref{eq: two-point correlation function of the memory}] of the outcomes vanish due to the feedback action.
However, one can compute the two-point correlation function of the detected outcomes in the steady state for the dynamics without feedback. In this case, the detections become correlated. Figure~\ref{fig: corr function no-FB Rabi} shows the steady-state two-point correlation function for two measurements separated by an interval $\delta t$ [$F_{\text{ss}}(\delta t)$] and $2\delta t$ [$F_{\text{ss}}(2\delta t)$].
Note that for a large interval between measurements, or equivalently $\lambda \delta t \gg 1$, the measurements become uncorrelated, as the two-point correlation function vanishes in this limit.

Let us consider the limit of fast measurements, $\delta t \ll \gamma,\lambda$, where the time interval $\delta t$ between successive measurements becomes infinitesimal. Expanding $P(x_n)$ in a Taylor series around $\delta t \approx 0$ (for arbitrary temperature $T$), we obtain
\begin{eqnarray}
    \label{eq: p_g in the FB ss for fast detections}
    P(x_n = -1) &=& 1 - \bar{N} \gamma\, \delta t \\
    &+& \frac{1}{2}(-2+\bar{N}(1+2\bar{N})~ p^2)\lambda^2~ \delta t ^2+ \mathcal{O}(\delta t^3),\nonumber
\end{eqnarray}
where $P(x_n = +1) = 1-P(x_n = -1)$, and $n\geq 2$. 
In the continuous monitoring limit $\delta t \rightarrow 0$, we have $P(x_n = -1) \rightarrow 1$, indicating that the feedback protocol effectively traps the system in the ground state, thereby cooling the qubit. 
Equation~\eqref{eq: p_g in the FB ss for fast detections} quantifies how the measurement interval influences the cooling performance of the protocol.
In other words, the smaller the measurement interval $\delta t$, the more effective the cooling in the feedback steady-state regime. 
This result also shows that, at high temperatures, fast measurements are essential for the feedback to efficiently cool the system.
Note that the measurement alone is insufficient to trap the system in the ground state; the feedback action, implemented through the rotations, is the fundamental mechanism that drives the system back to the ground state after each detection.
Finally, the average and variance of the detected outcomes for $n\geq2$ are given by
\begin{eqnarray}
        \braket{x_n} &=& -1 + 2\bar{N}\gamma \delta t\\
    && +(2-\bar{N}(1+2\bar{N})~ p^2)\lambda^2~ \delta t ^2+ \mathcal{O}(\delta t^3)\nonumber,
\end{eqnarray}
\begin{eqnarray}
    \text{Var}(x_n) &=& 4\bar{N}\gamma \delta t \\
    &&+2 (2-\bar{N}(1+4\bar{N})~ p^2)\lambda^2~ \delta t ^2+\mathcal{O}(\delta t^3)\nonumber,
\end{eqnarray}
where $p \equiv \gamma/\lambda$ represents the competition between the drive strength $\lambda$ and thermal coupling $\gamma$.

\section{Conclusion}
\label{sec:conclusion}

In this work, we have explored the properties of a general memory function that may be subject to arbitrary feedback strategies. 
We began by showing that the memory function can be interpreted as a classical system whose dynamics is coupled to that of the monitored quantum system. 
As a result, a general feedback dynamics can be described by an extended hybrid classical--quantum system, where the classical component accounts for the stochastic measurement outcomes (the memory), and the quantum component represents the monitored physical system. Within this framework, the feedback dynamics is naturally expressed as the evolution of a bipartite classical--quantum state.
This new representation of a general feedback strategy allows us to introduce informational measures that quantify the correlations between the classical memory and the quantum system. These measures provide deeper insight into the information exchange induced by the feedback process and offer direct applications at the interface between feedback control and quantum information science.

In addition, we introduced a general framework that provides the memory statistics, which can be applied both to feedback strategies based on the monitoring of quantum systems and to monitored systems without feedback actions.
Our framework is based on the evolution of the memory-resolved state, which fully characterizes the statistics of the stochastic memory, including its moments, time correlations, and steady-state distributions.
We provide illustrative examples of memory functions that are commonly used in feedback protocols. First, we considered the context of quantum jump monitoring, presenting two memory functions that record the last jump channel and the time elapsed since its detection. The second scenario corresponds to the case where the memory consists of the presently detected outcome, $y_t = x_t$, so that the memory statistics directly reflect the statistics of the measurement outcomes.

Finally, we present two feedback protocols where we apply the formalism developed for the memory function. 
The first case corresponds to a feedback protocol that stabilizes the qubit in the excited state against the interaction with a thermal bath, based on quantum jump detections. 
In this case, we discuss the jump-memory probability distribution, as well as its correlations with the system and other observables.
In the second example, we present a feedback protocol, previously implemented experimentally, that stabilizes the Rabi oscillations of a qubit against thermal decay using projective measurements. In this protocol, the feedback is based solely on the presently detected outcome, and we discuss the correlations and probability distribution of the measurement outcomes.
Our results provide a unified theoretical framework that not only encompasses earlier experimental demonstrations but also enables the design and analysis of more advanced feedback strategies.

Several questions remain open for future work. For instance, the dynamics considered here can be viewed as a sequential application of linear maps (instruments) to a quantum system, in close analogy with the process-tensor framework~\cite{PRXQuantum.2.030201,PhysRevLett.134.200401}. In our formalism, feedback control is incorporated by allowing the instruments to depend on the previous measurement record encoded in the memory. Exploring the connection between our formalism and the process-tensor approach is an interesting direction for future research and is currently the subject of ongoing work.

\vspace{5mm}

\begin{acknowledgments}
P.P.P. acknowledges funding from the Swiss National Science Foundation (Eccellenza Professorial Fellowship PCEFP2\_194268).
\end{acknowledgments}

\bibliography{feedback-refs}

\end{document}